\documentclass[final,3p]{elsarticle}

\usepackage{natbib}

\usepackage{graphicx,pstricks}

\usepackage{amsmath,amssymb,amsfonts,mathrsfs,esint,cancel,latexsym,mathtools,bm,scalerel,nicematrix,empheq,soul}

\newtheorem{lemma}{Lemma}
\newtheorem{theorem}{Theorem}
\newtheorem{remark}{Remark}

\newproof{pf}{Proof}
\newtheorem{definition}{Definition}
\newtheorem{corollary}[theorem]{Corollary}
\newtheorem{assumption}[theorem]{Assumption}

\usepackage{graphicx,pstricks}
\usepackage{amsmath,amssymb,amsfonts,mathrsfs,esint,cancel,latexsym,mathtools,bm,scalerel,nicematrix,empheq,soul}

\newcommand{\mean}[1]{\mathcal{E}\left\{ #1 \right\}}

\begin{document}

\begin{frontmatter}

\title{Mean square stability conditions for platoons with lossy inter-vehicle communication channels\tnoteref{footnoteinfo}}

\tnotetext[footnoteinfo]{This work was supported by the Chilean National Agency for Research and Development (ANID) through the scholarship program ``Doctorado Nacional/2020-21202404" and Fondecyt Iniciación Project 11221365. The material in this paper was not presented at any conference. Corresponding author F.J. Vargas.}

\author[First]{Marco A. Gordon}\ead{marco.gordon@sansano.usm.cl}   
\author[First]{Francisco J. Vargas}\ead{francisco.vargasp@usm.cl}            
\author[Second]{Andrés A. Peters}\ead{andres.peters@uai.cl}

\address[First]{Electronic Engineering Department, Universidad Técnica Federico Santa María, 2390123, Valparaíso, Chile}
\address[Second]{Faculty of Engineering and Sciences, Universidad Adolfo Ib\'a\~nez, Santiago 7941169, Chile}

\begin{abstract}
This paper studies the mean-square stability of heterogeneous LTI vehicular platoons with inter-vehicle communication channels affected by random data loss. We consider a discrete-time platoon system with predecessor following topology and constant time-headway spacing policy. Lossy channels are modeled by Bernoulli processes allowed to be correlated in space. We make use of a class of compensation strategies to reduce the effect of data loss. Necessary and sufficient conditions are derived to guarantee the convergence of the mean and variance of the tracking errors, which depend not only on the controller design but also on the compensation strategy and the probabilities of successful transmission. Through numerical simulations, we illustrate the theoretical results, describing different platoon behaviors. We also provide insights on the mean-square stability as a necessary condition for string stability in this stochastic setting. 
\end{abstract}

\begin{keyword}                           
Mean square stability; vehicle platoons; lossy channels.
\end{keyword}

\end{frontmatter}

\section{Introduction}
Vehicular platooning, where a set of vehicles travel at a cruise velocity, while maintaining a desired inter-vehicle distance, is a research topic with potential for achieving increased safety and better performance in roads \cite{turri2016cooperative,li2017platoon,studli2017vehicular}. A simple case considers that each vehicle follows its immediate predecessor. An independent leader, usually following an arbitrary trajectory of constant speed, is located at the front forming a string. The leading agent then determines the global trajectory and speed for the collection of vehicles. When the platoon is assumed to be in a networked environment, where agents exchange information through wireless channels, the study is extended to the field of Cooperative Adaptive Cruise Control (CACC) \cite{naus2010string,wang2018review}. CACC may be able to better enhance the safety and efficiency of road networks \cite{xu2014communi,thormann2020safe}.
Platoons possess several characteristics. For instance, the dynamical model of the vehicles can be assumed to be linear or non-linear, or defined in continuous or discrete-time \cite{middleton2010string,wen2018cooperative,luo2021leader}. If all the vehicles have the same dynamic model, the platoon is called homogeneous \cite{ploeg2014graceful,elahi2021h,middleton2010string}; otherwise, it is called heterogeneous \cite{naus2010string}.
The desired inter-vehicle distance can be constant, or variable. The employment of a constant time-headway spacing policy increases the inter-vehicle spacing as the speeds of the vehicles increase \cite{zhao2020stability,ploeg2013controller,qin2016stability,peters2016cyclic}.
The information flow topology of a platoon specifies how the information is exchanged from vehicle to vehicle. A commonly used topology is predecessor following (PF), where the communication is unidirectional and occurs only between two consecutive vehicles \cite{li2017platoon}. Most of the literature considers that the vehicle-to-vehicle communication channels are not affected by random issues. However, in reality, wireless channels are subject to problems such as random delays, noise, random data loss, etc. In such cases, the platoon becomes stochastic due to the nature of the communication channels.  
In this paper, we focus on platooning with lossy inter-vehicle communication channels. 

The performance of a platoon is analyzed through the tracking error that measures both platoon tasks, position tracking, and maintaining the inter-vehicle distance under constant speeds in steady state. The platoon tracking errors then must converge in time as a primary objective. However, due to the concatenation of vehicles, it is also important to ensure that disturbances do not amplify along the string. This property is known as string stability and guarantees the convergence and scalability of the platoon \cite{seiler2004disturbance,peppard1974string,liu2021stability}. 
For stochastic platoon setups, the string stability analysis is currently incipient \cite{feng2019string}, and presents a lack for a consistent formal definition of stochastic string stability. However, some works present an approach based on the statistics of the signal of interest (see for instance \cite{vegamoor2021string, qin2016stability,acciani2021stochastic,vargas2018string,gordon2021comparison,gordon2020platoon}). It is natural then, as an initial step, to focus on studying the convergence properties of the platoon statistics. 
In particular, the analysis of the convergence in time can be done through the notion of mean square stability (MSS) \cite{li2021event}. 


MSS has been widely studied in the field of Networked Control Systems (NCS) for several types of communication constraints, such as channels subject to signal to noise ratio constraints, random delays, multiplicative noise and data dropouts, among others \cite{vargas2013stabilization,gonzalez2019mean,schenato2007foundations, maass2016optimal,pang2019}. In multi-agent systems (MAS) applications, MSS has also been studied, mostly associated with the notion of mean-square consensus \cite{gong2017novel,xu2018mean,zheng2019consensusability}.
In the field of vehicular platooning, MSS literature is less extensive. 
In \cite{qin2016stability}, platoons with stochastic delays induced by data dropouts are studied. Sufficient conditions for stability are obtained by analyzing the mean and covariance dynamics. 
{The authors of \mbox{\cite{tang2018consensus}} obtain sufficient conditions for mean-square consensus based on LMIs  with a  topology more complex than PF but considering some specific type of homogeneous platoons.}
Platoons with communication affected by additive noise are considered in \cite{gordon2020platoon}, where necessary and sufficient conditions were obtained for mean square convergence.
In \cite{zhao2020stability}, topologies where the information of multiple predecessors are received by a follower are considered, and sufficient conditions for MSS are provided for heterogeneous time-independent random packet loss. 
Platoons with communication delays and packet dropouts are considered in \cite{elahi2021h}, where sufficient conditions for mean square consensus stability are derived.
In \cite{acciani2021stochastic}, the authors propose a controller designed to guarantee the convergence of the expected trajectory and minimize its variance in a platoon with lossy channels.

In a platooning setting, when dealing with random packet loss in a communication channel, the local controller in the closed-loop of each vehicle should not wait for the lost data to make a decision. In practice, it is required to use a policy or a data-loss compensation strategy to deal with the missing information \cite{li2019string}. 
Common strategies in the field of NCS consider replacing the missing information, for instance, with an estimate \cite{maass2016optimal}, or with previous available data \cite{acciani2021stochastic} (to-hold strategy), or to set the signal to zero (to-zero strategy) \cite{schenato2009zero}. In the context of platooning with lossy channels, most of the available works use one specific compensation strategy. For instance, \cite{tang2018consensus,wen2018cooperative,zhao2020stability,qin2016stability} use the last available measurement when the inter-vehicle communication fails. In \cite{vargas2018string} and \cite{elahi2021h}, a to-zero type protocol was applied to the controller input and the plant input respectively. 
In \cite{gordon2021comparison}, a numerical study was performed  with several types of data-loss compensation strategies adapted to a platooning setup, and it was observed that the convergence of the tracking error statistics depends on the chosen strategy, although these results lack an analytical explanation. In the present paper, we analyze the MSS of heterogeneous platoons of vehicles modeled as linear discrete-time systems with PF topology and a constant time-headway spacing policy. Inter-vehicle communication is considered to be affected by lossy channels, which are modeled as spatially correlated Bernoulli processes. 
The main contributions of this work are as follows:
\begin{itemize}
\item We obtain necessary and sufficient conditions for a platoon to be mean square stable, providing a characterization of the mean and variance of the tracking errors in the platoon under analysis, and studying their convergence properties.
Such conditions depend not only on the controller design and time-headway policy, but also on the channels arrival rates and on the compensation strategy. More precise conditions are derived for the case with independent lossy channels.
 \item Our analysis is valid for vehicles having different dynamical models, controllers, time headway values, and compensations strategies. Additionally, the inter-vehicle communication channels are also heterogeneous, and can be spatially correlated. The  case of independent channels for both heterogeneous and  homogeneous platoons are studied as a special cases.
 \item We consider not only one, but a wide class of compensation strategies to deal with missing data. We provide explicit conditions on the transfer function resulting from including such strategies, that ensure the platoon is MSS compatible. Our results also show that the chosen data-loss strategy plays a crucial role for MSS. This provides an analytical explanation to the performance of the strategies analyzed in \cite{gordon2021comparison}. 
 
  
  \item  We present a set of numerical examples to illustrate our results,  support our findings and enhance the related discussion.
   Here we also discuss about the implication of MSS on future research on stochastic string stability. 
\end{itemize}

The paper has the following outline: Section 2 presents the platoon setup under analysis. In Section 3 we propose a general state-space representation of the platoon. Section 4 presents necessary and sufficient conditions for MSS. In Section 5 we show numerical simulations and discuss about the implication of our results on string stabilization problems. Finally, conclusions and future work are discussed in Section 6.

\subsection{Notation and preliminaries} \label{sec:notation}
Let $M \in \mathbb{R}^{m\times m}$ be a real square matrix with eigenvalues $\lambda_1,...,\lambda_m$. We use $\rho(M) \triangleq \textrm{max} \{ |\lambda_1|,...,|\lambda_m| \}$ to denote the spectral radius of $M$. We will refer to $M$ as stable if $\rho(M)<1$. Given three matrices $A \in \mathbb{R}^{p\times q}$, $B \in \mathbb{R}^{q\times r}$ and $C \in \mathbb{R}^{r\times s}$, 
the following property holds \cite{bernstein2009matrix}
\begin{equation}
    \label{ec:prop_kronecker}
    vec(ABC) = (C^\top \otimes A)\ vec(B),
\end{equation}
where $vec$ denotes the vectorization of a matrix and $\otimes$ represents the Kronecker product. Let $Y \in \mathbb{R}^{m\times n}$ and $y \in \mathbb{R}^{mn\times 1}$ such that $vec(Y)=y$, the inverse of the $vec$ operator is such that $vec^{-1}(vec(Y))=Y$. 
Let $z \in \mathbb{R}^{l \times 1}$ be a column vector of length $l$, $M \in \mathbb{R}^{l\times l} ,\  N \in \mathbb{R}^{l\times l}$ and $D_z \in \mathbb{R}^{l\times l}$ a diagonal matrix $D_z=diag(z)$, then
\begin{align}
    \label{ec:prop_schur_def}
    D_z M {D_z}^{\top} &= (zz^{\top}) \odot M, \\
    \label{ec:prop_schur_diag}
    vec(M \odot N) &= diag(vec(M)) vec(N)
\end{align}
hold \cite{bernstein2009matrix}, where $\odot$ denotes the Hadamard product.
Let $x(k)$ and $y(k)$ be random variables of a stochastic process at the discrete time $k \in \mathbb{Z}$. The mean $\mu_x(k)$, covariance $P_x(k)$, and cross covariance $P_{xy}(k)$ are defined as $\mu_x(k) \triangleq \mean{x(k)}, \ P_x(k) \triangleq \mean{\bar{x}(k) \bar{x}(k)^\top}, \ P_{xy}(k) \triangleq \mean{\bar{x}(k) \bar{y}(k)^\top}$, 
where $\mean{\cdot}$ denotes the expectation operator and 
$\bar{x}(k) \triangleq x(k)-\mean{x(k)}.$
\section{Platooning Setup}\label{sec:setup}
This section presents the characteristics of the platooning setup under consideration for the analysis.
\subsection{Platoon with ideal communication}
We consider an interconnection of autonomous vehicles that consists of a leading vehicle and $N \in \mathbb{N}$ followers, all modeled as discrete-time linear time-invariant (LTI) systems. 
Each vehicle is identified with labels $i=0,\ldots,N$ that refer to their place within the string; $i=0$ represents the leader. The collection of vehicles is heterogeneous, having a plant model $G_i(z)$ and a local controller $K_i(z)$. The platoon is then controlled in a decentralized fashion, that is, each vehicle has access to limited information, and the control of each agent is designed locally. The platoon is assumed to have wireless communication capabilities in each vehicle and to move in a straight line with a predecessor-following topology. This configuration means that the vehicle-to-vehicle (V2V) communication is unidirectional and occurs only between an agent and its nearest follower. We consider that through the wireless channel, each vehicle has access to the position of its predecessor. 

\begin{figure}[!t]
    \begin{center}
    \includegraphics[width=0.92\columnwidth]{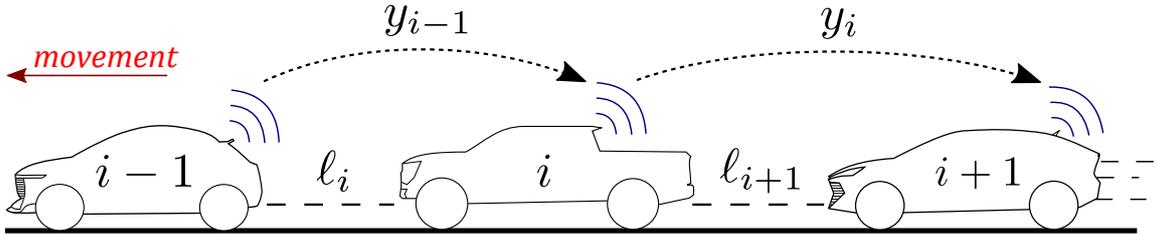}
    \caption{Platoon formation of three consecutive vehicles}
    \label{fig:platoon}                     
    \end{center}                             
\end{figure}

\begin{figure}[!t]
    \begin{center}
    \includegraphics[width=0.73\columnwidth]{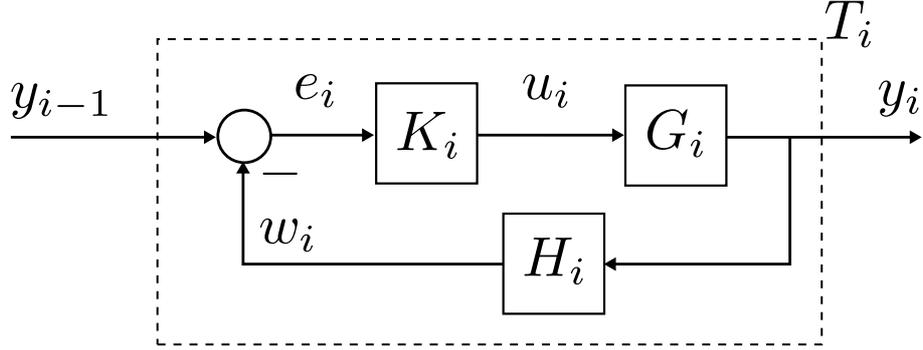}
    \caption{Closed loop of the $i$-th vehicle}
    \label{fig:closed_loop}        
    \end{center}                     
\end{figure}

We use $y_i(k)$  to denote the position of the $i$-th agent at the discrete time instant $k$. For simplicity in the exposition, vehicles will be considered to have zero length \cite{oncu2012string}, and thus $y_i(k)$ represents the position of a point in the real line. This assumption has no impact on the controller design nor on the subsequent stability analysis \cite{ploeg2014graceful,xu2014communi}. The inter-vehicle distance $\ell_i(k)$ between the $i$-th agent and its predecessor is given by $\ell_i(k)=y_{i-1}(k)-y_i(k)$. An illustrative scheme of the platoon  is given in Fig. \ref{fig:platoon}. 
The performance signal is the tracking error, denoted as $\zeta_i$. It is defined as the difference between the measured inter-vehicle distance $\ell_i$ and the desired separation $r_i$
\begin{equation}
    \zeta_i(k)=\ell_i(k)-r_i(k).
\end{equation}
In platooning designing it is common to consider configurations that achieve string stability, which is an important property for this type of application \cite{wang2018review,li2017platoon}. A platoon with PF topology can be compatible with string stability by implementing a 2-dof architecture \cite{naus2010string}. This requirement is commonly achieved by modifying the desired inter-vehicle distance. For instance, the constant time-headway spacing policy increases the separation between two consecutive vehicles according to the speed at which the $i$-th vehicle travels \cite{ploeg2014graceful,swaroop1994comparision,zhao2020stability,peters2014leader}. With this in mind, the desired inter-vehicle distance is set as  
\begin{equation}
    \label{ec:desired_separation}
    r_i(k)=\epsilon_i + h_i(y_i(k)-y_i(k-1))
\end{equation}
where $\epsilon_i$ is the safety distance and $h_i>0$ is a value known as the time-headway constant that weighs the measured discrete-time representation of the speed, i.e. $\nu_i(k) = y_i(k)-y_i(k-1)$. Notice that $\nu_i(0)$ is the initial speed of the $i$-th vehicle. Starting from rest we have that $y_i(-1)=y_i(0)$. Without loss of generality, the safety distance is set to zero for simplicity \cite{acciani2021stochastic,naus2010string}. Consequently, the tracking error is calculated as
\begin{equation}
    \label{ec:tracking_error}
    \zeta_i(k)=y_{i-1}(k)-w_i(k)
\end{equation}
where $w_i(k)=(h_i+1)y_i(k)-h_i\ y_i(k-1)$. This error measures the performance of both platoon tasks, namely, maintaining the separation distance, and tracking the leader position.
The control loop of one vehicle is shown in Fig. \ref{fig:closed_loop} where the vehicle dynamic of the $i$-th agent is given by $G_i(z)$.
The transfer function $H_i(z)$ incorporates the time-headway constant and it is defined as 
\begin{equation}
\label{eq:H}    
H_i(z)=(1+h_i)-h_i\ {z}^{-1}.
\end{equation}
Note that the output of $H_i$ corresponds to the feedback signal $w_i(k)$ introduced in \eqref{ec:tracking_error}. We use $u_i(k)$ to denote the output of the local controller $K_i$. 
In Fig. \ref{fig:closed_loop}, $e_i(k)$ represents the local control error, which is equal to the tracking error $\zeta_i(k)$ if the communication is perfect. In the next subsection, the distinction between these two types of errors becomes relevant.
For the given two degree-of-freedom architecture, the complementary sensitivity function $T_i(z)$ is 
\begin{align}\label{eq:T}
	T_i(z)=\frac{G_i(z)K_i(z)}{1+G_i(z)H_i(z)K_i(z)}.
\end{align}
Note that we are not interested in designing the controllers $K_i$, nor the time-headway $h_i$. Instead, we assume that they are properly designed in an ideal communication scenario and focus on analyzing the effect that lossy channels may have on the platoon. 
\begin{assumption}
    \label{ass:transfer_function}
      For $i=\{0,1,\ldots,N\}$ we assume that:
    \begin{enumerate}
    \item $T_i(z)$ is stable and strictly proper.
    \item $T_i(z)$ has no unstable cancellations in the product $G_i(z)H_i(z)K_i(z)$.
    \item The position of the leader $y_0(k)$ converges to a ramp signal.
    \item  The product $K_i(z)G_i(z)$ must have at least double integral action (two poles at $z=1$).
    \end{enumerate}
\end{assumption}
The first two are standard for the design of a closed loop system $T_i(z)$.
The third one is made to consider a platoon moving at a cruise velocity in steady-state.
This requirement does not remove the existence of disturbance changes in the speed (acceleration and breaking). 
The last assumption provides zero steady-state tracking error with a ramp reference and can be easily guaranteed through the controller design.

\subsection{Platooning with data-loss}
The transfer function between the output of the $(i-1)$-th agent and the output of its immediate follower ($i$-th agent), corresponds to $T_i(z)$ in the case of ideal inter-vehicle communication channels. However, when random data loss is considered, the signals involved may be stochastic, and the model of the  communication channels affects the interconnected system \cite{seiler2005h}.  The use of policies or compensation strategies to deal with data-loss becomes unavoidable, and may modify the aforementioned transfer function \cite{wen2018cooperative}.
The trajectories of the vehicles are not expected to be smooth in this scenario due to the effect of the random losses. A particular realization may exhibit erratic behavior and underperform in a practical sense. However, the effect of the unreliable channels may be mitigated by designing the dynamical features of the mean and variance of the tracking error.

\begin{figure}[!t]
    \begin{center}
    \includegraphics[width=0.95\columnwidth]{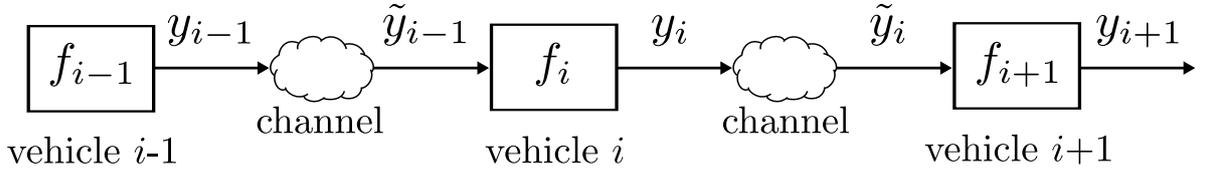}
    \caption{Scheme of the platoon interconnected through lossy channels}
    \label{fig:scheme_string}                     
    \end{center}                             
\end{figure}

Fig. \ref{fig:scheme_string} presents a general scheme of a platoon with lossy inter-vehicle communication channels where $f_i(\cdot)$ denotes the resulting feedback loop function after including a data-loss compensation scheme. The effect of the adopted strategies on the definition of $f_i(\cdot)$ is discussed in the next subsection. 
\begin{figure*}[t]
    \begin{center}
    \includegraphics[width=0.87\textwidth]{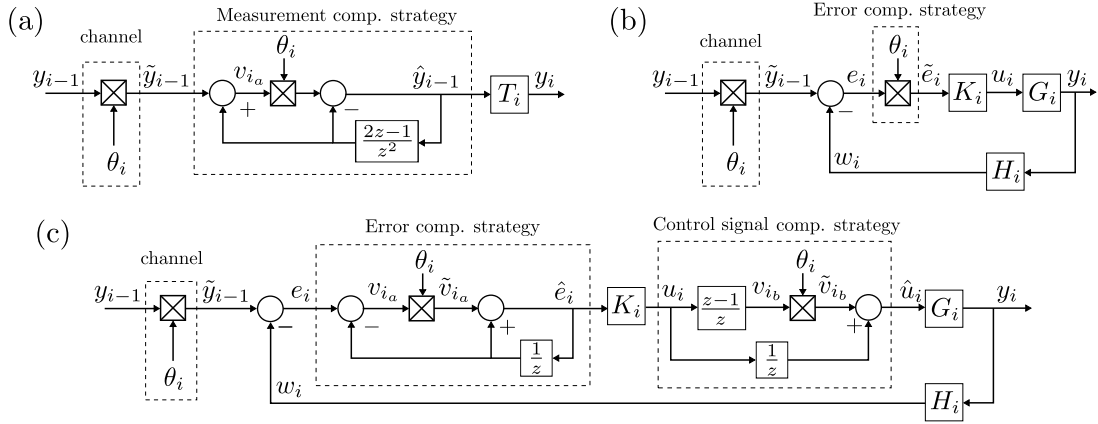}
    \caption{Block diagram of a vehicle with (a) measurement compensation strategy, (b) error compensation strategy, (c) mixed strategies (error and control signal based).}
    \label{fig:strat_example}        
    \end{center}   
\end{figure*}
Lossy channels are modeled as erasure channels defined by a Bernoulli stochastic process $\theta_i \in \{ 1,0 \}$ that describes whether the transmitted data is received or lost. When $\theta_i(k)=1$, the data is received successfully. Conversely, when $\theta_i(k)=0$ the data is considered lost (not available). The channel output is denoted as $\tilde{y}_{i-1}$ with
\begin{equation}
    \label{ec:channel_output}
    \tilde{y}_{i-1}(k) = \theta_i(k) y_{i-1}(k).
\end{equation}
\begin{assumption}
\label{assu:channel}
    We assume that each individual lossy channel $\theta_i$, with $i=1,2,\dots, N$ is an independent and identically distributed processes (i.i.d) with successful communication probability $p_i$. Thus, the mean $\mu_{\theta_i}$, and variance, $P_{\theta_i}$, of each channel are constant, and given by
    \begin{align}
        \mu_{\theta_i}=p_i, \qquad \qquad 
        P_{\theta_i}=p_i(1-p_i).
    \end{align}
    Additionally, we assume that $\theta_{i}(k)$ could be correlated with $\theta_{j}(k)$, for $i \neq j$. Thus, the covariance  $P_{\Theta}$ of the vector $\Theta(k) := \left[ \theta_1(k), \cdots,\theta_N(k) \right]^{\top}$
 could be non diagonal.
\end{assumption}
Assumption 2 implies that the channels are uncorrelated in time but possibly correlated in space. Thus, there could be a statistical link between two pairs of channels given by $\theta_i$ and $\theta_j$. The case where channels are spatially independent is given as a special case in Section \ref{sec:MSS_conditions}. Clearly, the value of $p_i$ depends on several factors, but strongly on the transmitting and receiving devices of the $(i-1)$-th  and $i$-th vehicles respectively.

The local error $e_i$ should not be confused with the tracking error $\zeta_i$. Indeed, only for ideal communication, both errors are equivalent. Given that $e_i$ utilizes the channel output signal \eqref{ec:channel_output}, it does not measure the true inter-vehicle distance error. Thus, it is suitable to use $\zeta_i$ (see \eqref{ec:tracking_error}) as the performance signal in this setting, since it is calculated using the true position of the vehicles. 

\subsubsection{Strategies to deal with data-loss}
A compensation strategy refers to the policy or action that is taken by the receiver when the information transmitted by the predecessor does not arrive.
Commonly used strategies either hold the previous value of a signal or replace it with zero when dropouts occur \cite{li2019string}. Besides, more than one strategy can be used simultaneously and be applied to the measured position, the local error, or the control signal \cite{gordon2021comparison}. The use of an appropriate compensation strategy may have a positive impact on the MSS and also on the string stability of the platoon \cite{gordon2021comparison}. Said strategies modify the closed-loop structure and the resulting state-space representation of the interconnected system.
To illustrate this, Fig. \ref{fig:strat_example} presents three types of control strategies applied to different signals of the control loop. In Fig. \ref{fig:strat_example}$(a)$ a measurement compensation strategy is included after the channel output to estimate the position. This strategy is described as
\begin{equation}
    \hat{y}_{i-1}(k)= \tilde y_{i-1}(k)-\eta_i(k)\theta_i(k) + \eta_i(k).
\end{equation}
where $\eta_i(k)=2\hat y_{i-1}(k-1)-\hat y_{i-1}(k-2)$ is the estimated distance traveled in a lapse of time between $k-1$ and $k$. 
An error compensation strategy like the one in Fig. \ref{fig:strat_example}$(b)$ sets the error to zero when the position is lost. This type of strategy is described as
\begin{equation}
    \label{ec:error_anterior}
    \hat{e}_{i}(k) = e_i(k) \theta_i(k)
\end{equation}
Fig. \ref{fig:strat_example}$(c)$ shows a combination of strategies. 
The error is replaced with its previous value \cite{vargas2018string}, and the control signal is replaced with its previous estimated value in events of data loss. These strategies are described as
\begin{align}
    \label{ec:error_zero}
    \hat{e}_{i}(k) &= \left[ e_{i}(k) - \hat e_{i}(k-1) \right] \theta_i(k) + \hat e_{i}(k-1), \\
    \hat{u}_{i}(k) &= \left[ u_{i}(k) - u_{i}(k-1) \right] \theta_i(k) + u_{i}(k-1)
\end{align}

For all compensation strategies, their initial conditions must be specified with a given policy, especially those depending on past values. For instance, we can set $\hat{e}_i(-1)=0$ and $\hat{u}_i(-1)=0$ for the strategy in Fig. \ref{fig:strat_example}$(c)$. 
The above strategies are particular examples, but our analysis is not restricted to them.
The structure of the strategies directly affect the resulting functions $f_i(\cdot)$ that relate $\tilde y_{i-1}(k)$ with $y_{i}(k)$. This motivates the use of a general representation of the closed-loop system. 
\begin{assumption}
\label{assu:strategies}
    We consider strategies that can be written as linear feedback loops subject to multiplicative Bernoulli channels, such as those in Fig. \ref{fig:strat_example}. 
\end{assumption}
\subsubsection{Problem Statement}
In the present paper, we study the time-convergence of the platoon. We first consider that each closed loop $T_i$ is properly designed and internally stabilizes the platoon assuming ideal communication. Then, we assume lossy communication in each inter-vehicular channel and analyze the stability of the platoon.
The stability analysis of linear systems with random data loss can be performed through the property of mean square stability which guarantees the convergence of the mean and variance of the signal of interest. 
The signal of interest in our case is the tracking error vector defined as $\bm{\zeta}(k)=\begin{bmatrix}
    \zeta_1(k) &  \hdots & \zeta_N(k)
    \end{bmatrix}^{\top}.$

\begin{definition}
The platoon under analysis is mean square stable if and only if  the mean and covariance matrix of the tracking error vector $ \bm{\zeta}(k)$ are such that 
\begin{equation}
    \label{ec:MSS_def}
    \lim_{k \to \infty} \mu_{\bm{\zeta}}(k) = \mu_{\bm{\zeta}} \quad \text{and} \quad
    \lim_{k \to \infty} P_{\bm{\zeta}}(k) = P_{\bm{\zeta}},
\end{equation}
for some finite $\mu_{\bm{\zeta}}$ and $P_{\bm{\zeta}}.$ 
\end{definition}
Our interest is to determine  necessary and sufficient conditions for a platoon to be mean square stable. 


\begin{figure}[!t]
    \begin{center}
    \includegraphics[width=0.95\columnwidth]{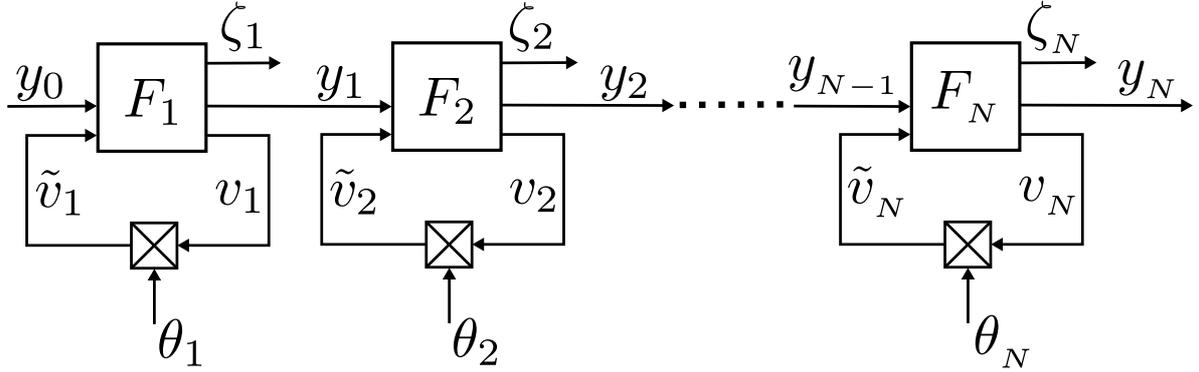}
    \caption{Alternative platoon representation }
    \label{fig:concatenacion}                
    \end{center}                             
\end{figure}

\section{State-space representation}\label{sec:SS_representation}

Given Assumption \ref{assu:strategies}, the platoon can be represented as depicted in Fig. \ref{fig:concatenacion} where the lossy channels $\theta_i$ are separated  from  the remaining linear system $F_i$, which is defined by $G_i$, $K_i$, $H_i$ and the applied data-loss compensation mechanism. 
Hence, a platoon with PF topology can be viewed as a cascade interconnection of systems, although there could be a statistical link among these systems beyond the cascade structure. It is convenient to use this representation as it is independent of the internal structure of $F_i$ and provides flexibility for the inclusion, for example, of several data compensation strategies (see for instance \cite{gordon2021comparison}).

In this representation each $F_i$, for $i\geq 1$, is assumed to have the following minimal state-space description
\begin{subequations}
\begin{align}
    \label{ec:state_single}
    x_i(k+1) &= A_i\ x_i(k) + B_i \tilde{v}_i(k), \quad x_i(0)={x_i}_0 \\
    \label{ec:position_single}
    y_i(k) &= C_{y_i}\ x_i(k) \\
    \label{ec:error_single}
    \zeta_i(k) &= C_{\zeta_i} \ x_i(k) + D_{\zeta_i}\ y_{i-1}(k) \\
    \label{ec:canal_single}
    v_i(k) &= C_{v_i} \ x_i(k) + D_{v_i} \ y_{i-1}(k)
\end{align}    
\end{subequations}
where $x_i(k) \in \mathbb{R}^{n_x}$ is the system state, $v_i(k) \in \mathbb{R}^{n_v}$ and $\tilde{v}_i(k) \in \mathbb{R}^{n_v}$ are the input and  the output of the $i$-th Bernoulli link, and $A_i$, $B_i$, $C_{y_i}$, $C_{\zeta_i}$, $C_{v_i}$, $D_{\zeta_i}$, $D_{v_i}$ are real matrices of appropriate dimensions. 
For some particular compensation strategies, $\zeta_i(k)$ could be equal to $v_i(k)$, but in general these are different signals. Indeed, $\zeta_i(k)$ is a scalar signal, while $v_i(k)$ could be not, e.g., when using strategy in Fig. \ref{fig:strat_example}$(c)$, where  $v_i(k)=[v_{i_a}(k) \,\,  v_{i_b}(k)]^{\top}$. 

\begin{remark}
Naturally, the behavior of the systems $F_i$ depends on the arrival rate $p_i$, however it can also depend on other channels statistics. For instance, we can define $\tau_i =  (p_i + p_{i-1}+ \cdots p_{i-\ell}) /\ell $ and use it as a measure of the communication quality of $\ell$ vehicles ahead to define, with a given criteria, the time headway $h_i$, the  controller gain, a  parameter of the compensation strategy, etc. \end{remark}

The concatenation of the systems $F_i$ can also be represented using a generalized feedback loop with a zero-mean multiplicative MIMO channel. Since the channels are considered temporally independent, we first define the matrix $\Theta_d$ and its mean $\Upsilon$ as follows:
\begin{align}
    \Theta_d(k) &= diag(\Theta(k))= diag(\theta_1(k), \cdots,\theta_N(k))\\
    \Upsilon &= diag(p_1,\dots,p_N).
\end{align}

\begin{figure}[!t]
    \begin{center}
    \includegraphics[width=0.27\columnwidth]{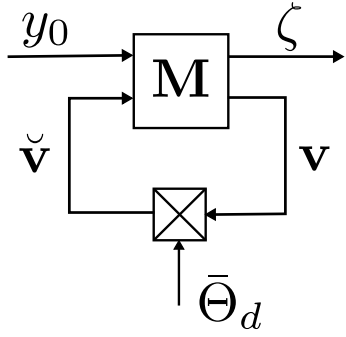}
    \caption{General feedback closed loop of the platoon system.}
    \label{fig:lazo_gen}                
    \end{center}                    
\end{figure}

We also define the zero-mean multiplicative channel matrix $\bar{\Theta}_d(k) = \Theta_d(k)-\Upsilon$. This allows us to write the platoon dynamics using the representation of Fig. \ref{fig:lazo_gen}, where $\mathbf{M}$ represents a linear time invariant system with state vector denoted by $\mathbf{x}$ and where $\mathbf{v}$ and $\breve{\mathbf{v}}$ represent the channel input and output respectively. The state and the channel input vectors are formed by the concatenation of the corresponding terms of each vehicle, that is,
\begin{align*}
    \mathbf{x}(k) &=\begin{bmatrix}
    x_1(k) & \hdots & x_N(k)
    \end{bmatrix}^{\top}, \\ 
    \mathbf{v}(k) &=\begin{bmatrix}
    v_1(k) & \hdots & v_{N}(k)
    \end{bmatrix}^{\top}, \;\, \breve{\mathbf{v}}=\bar{\Theta}_d(k)\mathbf{v}.
\end{align*}

\begin{lemma}\label{lema_ss}
The closed loop dynamics of the system $\mathbf{M}(z)$ can be written as
\begin{subequations}\label{ec:ss_lg}
\begin{align}
    \label{ec:estado_lg}
    \mathbf{x}(k+1) &= \mathbf{A}\ \mathbf{x}(k) + \mathbf{B}\ y_0(k) + \mathcal{B}  \breve{\mathbf{v}}(k) \\
    \label{ec:error_lg}
    \bm{\zeta}(k) &= \mathbf{C}_{\bm{\zeta}}\ \mathbf{x}(k) + \mathbf{D}_{\bm{\zeta}}\ y_0(k) \\
    \label{ec:entrada_canal_lg}
    \mathbf{v}(k) &= \mathbf{C_v}\ \mathbf{x}(k) + \mathbf{D_v} y_0(k)
\end{align}    
\end{subequations}
with the following state-space matrices:
\begin{align*}
    \mathbf{A} &=
    \begin{bmatrix}
        \alpha_1 & \ &  \  &   \  \\
        \gamma_2 & \alpha_2  &  \ &   \  \\
        \  & \ddots & \ddots & \ \\
        \ & \ & \gamma_N & \alpha_N \\
    \end{bmatrix},\;
    \mathbf{B} =
    \begin{bmatrix}
        p_1B_1D_{v_1} \\
        0 \\
        \vdots \ \\
        0 \\
    \end{bmatrix}, \\
    \mathcal{B}&= diag(B_1, \cdots B_N), \\
    \mathbf{C}_{\bm{\zeta}} &=
    \begin{bmatrix}
        C_{\zeta_1} & \ &  \  &   \  \\
        D_{\zeta_2}C_{y_1} & C_{\zeta_2}  &  \ &   \  \\
        \  & \ddots & \ddots & \ \\
        \ & \ & D_{\zeta_N}C_{y_{N-1}} & C_{\zeta_N} \\
    \end{bmatrix}, \ 
    \mathbf{D}_{\bm{\zeta}}=
    \begin{bmatrix}
        D_{\zeta_1} \\
        0 \\
        \vdots \ \\
        0 \\
    \end{bmatrix}, \\
    \mathbf{C_v} &=
    \begin{bmatrix}
        C_{v_1} & \ &  \  &   \  \\
        D_{v_2}C_{y_1} & C_{v_2}  &  \ &   \  \\
        \  & \ddots & \ddots & \ \\
        \ & \ & D_{v_N}C_{y_{N-1}} & C_{v_N} \\
    \end{bmatrix}, \ 
    \mathbf{D_v}=
    \begin{bmatrix}
        D_{v_1} \\
        0 \\
        \vdots \ \\
        0 \\
    \end{bmatrix}
\end{align*}
where $\alpha_i=A_i+p_iB_iC_{v_i}$ and $\gamma_i=p_iB_iD_{v_i}C_{y_{i-1}}$.

\end{lemma}
\begin{pf}
See appendix \ref{prf:lema_ss}.
\end{pf}
Given this representation, we can partition  $\mathbf{M}$ as follows
\begin{align}
    \begin{bmatrix}
    \bm{\zeta} \\ \mathbf{v}
\end{bmatrix} =
\underbrace{\begin{bmatrix}
   \mathbf{M}_{11} && \mathbf{M}_{12} \\  \mathbf{M}_{21} && \mathbf{M}_{22}
\end{bmatrix}}_{\mathbf{M}}\begin{bmatrix}
   y_{0} \\ \breve{\mathbf{v}}
\end{bmatrix},
\end{align}
and write the transfer function from $y_0$ to $\bm{\zeta}$ and $\mathbf{v}$ as
\begin{align} \nonumber
\mathbf{M}_{11}(z)&= \mathbf{C}_{\bm{\zeta}} (z\mathbf{I}-\mathbf{A})^{-1}\mathbf{B}+\mathbf{D}_{\bm{\zeta}}\\
\nonumber
\mathbf{M}_{21}(z)&= \mathbf{C}_{\mathbf{v}} (z\mathbf{I}-\mathbf{A})^{-1}\mathbf{B}+\mathbf{D}_{\mathbf{v}}.
\end{align}
\begin{remark}
As long as the platoon has an equivalent representation like in Fig. \ref{fig:lazo_gen} with a linear and time invariant $\mathbf{M}$, this general feedback loop scheme allows flexibility to analyze different platoon setups. For instance, $\mathbf{M}$ can describe homogeneous and heterogeneous platoons. This representation allows vehicles with different dynamic models, controllers, channel successful transmission probabilities, or compensation strategies. 
Note also that,
the current mathematical description may be suitable for describing interconnections of systems with local controllers, that may not necessarily represent vehicular systems. In that regard, the following results might have some use in multi-agent applications different from platooning.
\end{remark}

\section{MSS conditions} \label{sec:MSS_conditions}

In order to derive our main results, we first present in Lemma \ref{lema_statistics} below the first and second order statistics of the signals of interest.

\begin{lemma}\label{lema_statistics}
Consider a platoon described by the representation given in Lemma \ref{lema_ss}. The mean of the state, tracking error, and channel input, respectively, satisfy
\begin{subequations}\label{ec:media_lg}
    \begin{align}
    \label{ec:media_estado_lg}
    \mu_{\mathbf{x}}(k+1) &= \mathbf{A}\ \mu_{\mathbf{x}}(k) + \mathbf{B}\ y_0(k) \\
    \label{ec:media_error_lg}
    \mu_{\bm{\zeta}}(k) &= \mathbf{C}_{\bm{\zeta}}\ \mu_{\mathbf{x}}(k) + \mathbf{D}_{\bm{\zeta}}\ y_0(k) \\
    \label{ec:media_entrada_canal_lg}
    \mu_{\mathbf{v}}(k) &= \mathbf{C_v}\ \mu_{\mathbf{x}}(k) + \mathbf{D_v}\ y_0(k),
    \end{align}
\end{subequations}
and the corresponding covariance matrices satisfy
\begin{subequations}\label{ec:varianza_lg}
\begin{align}
    \label{ec:varianza_estado_lg}
    P_{\mathbf{x}}(k+1) &= \mathbf{A}\ P_{\mathbf{x}}(k)\  \mathbf{A}^\top + \mathcal{B} \left[ P_{\Theta} \odot P_{\mathbf{v}}(k) \right] {\mathcal{B}}^\top \notag \\ 
    &+ \mathcal{B} \left[ P_{\Theta} \odot (\mu_{\mathbf{v}}(k) \mu_{\mathbf{v}}(k)^\top )\right] {\mathcal{B}}^\top \\
    \label{ec:varianza_error_lg}
    P_{\bm{\zeta}}(k) &= \mathbf{C}_{\bm{\zeta}}\ P_{\mathbf{x}}(k)\ \mathbf{C}_{\bm{\zeta}}^\top \\
    \label{ec:varianza_entrada_canal_lg}
    P_{\mathbf{v}}(k) &= \mathbf{C_v}\ P_{\mathbf{x}}(k)\ {\mathbf{C_v}}^\top.
\end{align}    
\end{subequations}
\end{lemma}
\begin{pf}
See appendix \ref{pr:lemma_statistics}.
\end{pf}

The following theorem is the main result of this paper and presents necessary and sufficient conditions for MSS of platoons subject to random data-loss as the one described above.

\begin{theorem}
\label{theo:MSS}
Consider the platoon described in Section \ref{sec:setup}, with the state-space representation given in Section \ref{sec:SS_representation}. Then, 
\begin{enumerate}
    \item the mean 
$\mu_{\bm{\zeta}}(k)$ converges if and only if
\begin{equation}\label{eq:cond_mean}
\mathbf{M}_{11}(1)=0 \; \text{ and } \;      \rho(\mathbf{A})<1,   
\end{equation}
\item the variance $P_{\bm{\zeta}}(k)$ converges if and only if $ \rho(\mathbf{A})<1$ and
\begin{equation}\label{eq:cond_var}
\mathbf{M}_{21}(1)=0 \; \text{ and } \;      \rho((\mathbf{A} \otimes \mathbf{A})+\Delta)<1,   
\end{equation}
where $\Delta=(\mathcal{B} \otimes \mathcal{B}) \ diag(vec(P_{\Theta}))\ (\mathbf{C_v} \otimes \mathbf{C_v})$.
\end{enumerate}
Consequently,  the platoon is mean square stable if and only if  \eqref{eq:cond_mean} and \eqref{eq:cond_var} are met.
\end{theorem}
\begin{pf}
We define an alternative state $\mathbf{x}^*(k)=\mathbf{x}(k)-\mathbf{x}(k-1)$. From \eqref{ec:media_estado_lg} we have that
\begin{equation}
    \label{ec:media_estado_lg_ast}
    \mu_{\mathbf{x}^*}(k+1) = \mathbf{A}\ \mu_{\mathbf{x}^*}(k) + \mathbf{B}\ m_0(k), 
\end{equation}
where $m_0(k)=y_0(k)-y_0(k-1)$. Given Assumption \ref{ass:transfer_function}, the position input $y_0(k)$ converges to a ramp signal. Then,  $m_0(k)$ converges to a constant value which is the cruise velocity (slope of the position at instant $k$).
From \eqref{ec:media_error_lg} and \eqref{ec:media_entrada_canal_lg} we have 
\begin{align}
    \label{ec:media_error_lg_ast}
    \mu_{\bm{\zeta}}(k) &= \mu_{\bm{\zeta}}(k-1) + \mathbf{C}_{\bm{\zeta}}\ \mu_{\mathbf{x}^*}(k) + \mathbf{D}_{\bm{\zeta}}\ m_0(k)\\
     \label{ec:media_entrada_lg_ast}
    \mu_{\mathbf{v}}(k) &= \mu_{\mathbf{v}}(k-1) + \mathbf{C}_{\mathbf{v}}\ \mu_{\mathbf{x}^*}(k) + \mathbf{D}_{\mathbf{v}}\ m_0(k).
\end{align}
On the other hand, replacing \eqref{ec:varianza_entrada_canal_lg} in \eqref{ec:varianza_estado_lg}, and using the linearity property of the Hadamard product it follows that
\begin{align}
    \label{ec:mss_estado}
    P_{\mathbf{x}}(k+1) &= \mathbf{A} P_{\mathbf{x}}(k) \mathbf{A}^\top  + \mathcal{B} \left[ P_{\Theta} \odot (\mathbf{C_v} P_{\mathbf{x}}(k)\ {\mathbf{C_v}}^\top) \right] {\mathcal{B}}^\top \notag \\
    &\quad + \mathcal{B} \left[ P_{\Theta} \odot \left( \mu_{\mathbf{v}}(k)\ \mu_{\mathbf{v}}(k)^\top \right) \right] {\mathcal{B}}^\top.
\end{align}
Applying the $vec$ operator and property \eqref{ec:prop_kronecker} in \eqref{ec:mss_estado} yields
\begin{align*}
    \mathbf{X}(k+1) &= (\mathbf{A} \otimes \mathbf{A}) \mathbf{X}(k)  +(\mathcal{B} \otimes \mathcal{B})Z(k)+ S(k),
\end{align*}
where  $Z(k)=vec(P_{\Theta} \odot (\mathbf{C_v}\ P_{\mathbf{x}}(k)\ {\mathbf{C_v}}^\top))$, $S(k)= vec \left[ \mathcal{B} (P_{\Theta} \odot (\mu_{\mathbf{v}}(k) \mu_{\mathbf{v}}(k)^\top)) \mathcal{B}^\top\right]$ and $\mathbf{X}(k)=vec(P_{\mathbf{x}}(k))$. Using property \eqref{ec:prop_schur_diag}, we write
\begin{align*}
   Z(k)
    & = vec(P_{\Theta}) \odot vec(\mathbf{C_v}\ P_{\mathbf{x}}(k) {\mathbf{C_v}}^\top) \\
    &= diag(vec(P_{\Theta})) (\mathbf{C_v} \otimes \mathbf{C_v}) \mathbf{X}(k).
\end{align*}
Let $\Delta=(\mathcal{B} \otimes \mathcal{B}) diag(vec(P_{\Theta}))\ (\mathbf{C_v} \otimes \mathbf{C_v})$. Then, 
\begin{align}
    \label{ec:mss_varianza_estado_lg}
  \mathbf{X}(k+1) &= \left[ (\mathbf{A}\otimes\mathbf{A}) + \Delta  \right] \mathbf{X}(k) + S(k).
\end{align}
\textit{Necessity}: Consider the platoon MSS. Thus, we conclude from \eqref{ec:media_error_lg_ast} that there exists $\mu_{\bm{\zeta}}= \lim_{k \rightarrow \infty} \mu_{\bm{\zeta}}(k)$  satisfying  $\mu_{\bm{\zeta}} = \mu_{\bm{\zeta}} + \mathbf{C}_{\bm{\zeta}}\ \mu_{\mathbf{x}^*} + \mathbf{D}_{\bm{\zeta}}\ m_0$,
where $\mu_{\mathbf{x}^*}$ and $m_0$ 
are the stationary values of  $\mu_{\mathbf{x}^*}(k)$ and $m_0(k)$, respectively. This implies that $\mathbf{C}_{\bm{\zeta}}\ \mu_{\mathbf{x}^*} + \mathbf{D}_{\bm{\zeta}}\ m_0=0$.
From \eqref{ec:media_estado_lg_ast} we have  $\mu_{\mathbf{x}^*}= \mathbf{A}\ \mu_{\mathbf{x}^*} + \mathbf{B}\ m_0
    = (I-\mathbf{A})^{-1}\ \mathbf{B}\ m_0$, and thus  $ \mathbf{C}_{\bm{\zeta}}\ \mu_{\mathbf{x}^*} + \mathbf{D}_{\bm{\zeta}}\ m_0 = \left(\mathbf{C}_{\bm{\zeta}}(I-\mathbf{A})^{-1}\ \mathbf{B}+ \mathbf{D}_{\bm{\zeta}}\right) m_0=\mathbf{M}_{11}(1)m_0$. Hence,  MSS implies $\mathbf{M}_{11}(1)=0$. Also, it is clear that 
    $\mu_{\mathbf{x}^*} = (I-\mathbf{A})^{-1}\ \mathbf{B}\ m_0=\sum_{k=1}^{\infty} \mathbf{A}^{k} \mathbf{B} \ m_0$,
    which implies $\rho(\mathbf{A})<1$, otherwise the summation will not converge.
    This shows that \eqref{eq:cond_mean} is necessary for the mean convergence.
    
MSS assumption also implies that there exists  $P_{\bm{\zeta}}=\lim_{k \rightarrow \infty} P_{\bm{\zeta}}(k)$, which requires $P_{\mathbf{x}}(k)$ to converge to a constant matrix $P_{\mathbf{x}}$ satisfying
\begin{align}
    P_{\mathbf{x}} &= \mathbf{A} P_{\mathbf{x}} \mathbf{A}^\top 
     + \mathcal{B} \left[ P_{\Theta} \odot (\mathbf{C_v} P_{\mathbf{x}}\ {\mathbf{C_v}}^\top) \right] {\mathcal{B}}^\top \notag \\
    &\quad + \mathcal{B} \left[ P_{\Theta} \odot \left( \mu_{\mathbf{v}}\ \mu_{\mathbf{v}}^\top \right) \right] {\mathcal{B}}^\top,
\end{align}
where $\mu_{\mathbf{v}}$ is the stationary value of $\mu_{\mathbf{v}}(k)$. From \eqref{ec:media_entrada_lg_ast}, and mimicking the analysis above for $\mu_{\bm{\zeta}}$, it is easy to conclude that MSS implies $\mathbf{M}_{21}(1)=0$. 

Since $P_{\mathbf{x}}(k) \rightarrow P_{\mathbf{x}}$, then the vector 
$\mathbf{X}(k)$ converges too. It is clear from \eqref{ec:mss_varianza_estado_lg} that the stationary value $\mathbf{X}$ is given by 
$  \mathbf{X}= 
  (I-(\mathbf{A}\otimes\mathbf{A}) - \Delta)^{-1}\ S =\sum_{k=1}^{\infty} ((\mathbf{A}\otimes\mathbf{A}) + \Delta)^{k} S, $
where $S= vec \left[ \mathcal{B} (P_{\Theta} \odot (\mu_{\mathbf{v}} \mu_{\mathbf{v}}^\top)) \mathcal{B}^\top\right]$. We then conclude that if the platoon is MSS, then the spectral radius of $(\mathbf{A}\otimes\mathbf{A}) + \Delta$ is less than one. This proves \eqref{eq:cond_var} is necessary for the covariance convergence and thus for MSS.

\textit{Sufficiency}: Assume conditions in \eqref{eq:cond_mean} and \eqref{eq:cond_var}
are met. From \eqref{ec:media_estado_lg_ast} we have
\begin{equation}
\label{ec:sum_mu_ask}
    \mu_{\mathbf{x}^*}(k) = \mathbf{A}^k\ \mu_{\mathbf{x}^*}(0) +  \sum_{i=1}^{k}\mathbf{A}^{i-1}\mathbf{B}\ m_0(k-i).
\end{equation}
Since $\rho(\mathbf{A})<1$, then $\lim_{k \rightarrow \infty} \mathbf{A}^k=0$. It is known that $m_0=\lim_{k \rightarrow \infty} m_0(k) $, thus the summation in \eqref{ec:sum_mu_ask} converges and, hence, $\mu_{\mathbf{x}^*}(k)$ converges to a constant value $\mu_{\mathbf{x}^*}$. Such value can be written as $\mu_{\mathbf{x}^*}
    = (I-\mathbf{A})^{-1}\ \mathbf{B}\ m_0$. Thus, from \eqref{ec:media_error_lg_ast} we have that
    $ \lim_{k \rightarrow \infty} \left(\mu_{\bm{\zeta}}(k) - \mu_{\bm{\zeta}}(k-1) \right) =\mathbf{M}_{11}(1)m_0$. Since $\mathbf{M}_{11}(1)=0$, we conclude that $\mu_{\bm{\zeta}}(k)$ converges to a constant value $\mu_{\bm{\zeta}}$, proving that \eqref{eq:cond_mean} is sufficient for the mean convergence. Similarly, given that $\mathbf{M}_{21}(1)=0$, it is easy to see from \eqref{ec:media_entrada_lg_ast} that $\mu_{\mathbf{v}}(k)$ also converges. The latter implies that $S(k)$ converges to $S$. Finally, from \eqref{ec:mss_varianza_estado_lg} we have
    \begin{align}
\nonumber
    \mathbf{X}(k) = & \left[ (\mathbf{A}\otimes\mathbf{A}) + \Delta  \right]^k \mathbf{X}(0) \\
    \label{ec:sum_X_ask}
    &+  \sum_{i=1}^{k} \left[ (\mathbf{A}\otimes\mathbf{A}) + \Delta  \right]^{i-1} S(k-i). 
\end{align}
Since $\rho((\mathbf{A} \otimes \mathbf{A})+\Delta)<1$, and  $S(k)$ converges, we conclude that $\mathbf{X}(k)$ converges and so does $P_{\mathbf{x}}(k)$, which proves that both \eqref{eq:cond_mean} and \eqref{eq:cond_var} are sufficient for MSS. \qed
\end{pf}
Theorem \ref{theo:MSS} presents necessary and sufficient conditions for the platoons of interest to be stable in the mean square sense. Such conditions depend not only on the plant, controller dynamics, and time-headway, but also on the applied compensation strategy, the probabilities of successful transmission, and the correlation between channels. It is expected for the quality of the channels to have a relevant impact on the MSS of the platoon. Section \ref{sec:simulation} provides a simulation-based analysis about the influence that the probabilities $p_i$ may have on MSS for a two-follower platoon. 

The conditions for $\mathbf{M}_{11}(z)$ and $\mathbf{M}_{21}(z)$ in Theorem \ref{theo:MSS} are equivalent to requiring such systems to have at least one zero at $z=1$. $\mathbf{M}_{11}(z)$ and $\mathbf{M}_{21}(z)$ strongly depend on the chosen strategy to deal with data loss. It is important to note that, although $T(z)$ is required to have two zeros at $z=1$ to be able to track ramp signals, such zeros could not be part of $\mathbf{M}_{11}(z)$ and $\mathbf{M}_{21}(z)$, since the adopted protocol to deal with data loss could cancel one or both zeros. This reveals that the strategy to deal with data loss must be carefully chosen. The MSS convergence of the error could be to values different from zero for the mean and the variance. In the following corollary, we specify conditions to ensure that the mean and variance of the error converge to zero.
\begin{corollary}
\label{cor:conv_to_zero}
Consider a platoon satisfying the conditions in Theorem \ref{theo:MSS}. Then,
\begin{enumerate}
    \item  if $\mathbf{M}_{11}(z)$ has two or more zeros at $z=1$,  $\mu_{\bm{\zeta}}(k)$ converges to zero. Instead, if $\mathbf{M}_{11}(z)$ has only one zero at $z=1$, then $\mu_{\bm{\zeta}}(k)$ converges to a non-zero value given by
    \begin{equation}
        \mu_{\bm{\zeta}} = - m_0 {\mathbf{C}_{\bm{\zeta}}}(I -\mathbf{A})^{-2}\mathbf{B}, 
    \end{equation}
    \item  if $\mathbf{M}_{21}(z)$ has two or more zeros at $z=1$, then  $P_{\bm{\zeta}}(k)$ converges to zero. Instead, if $\mathbf{M}_{21}(z)$ has only one zero at $z=1$ then $P_{\bm{\zeta}}(k)$ converges to a non-zero value given by
    \begin{align}
  \label{ec:var_estacionaria}
        P_{\bm{\zeta}}=& \mathbf{C}_{\bm{\zeta}} \, vec^{-1}  \left[ \left( I-(\mathbf{A}\otimes\mathbf{A}) - \Delta \right)^{-1} S\right] \mathbf{C}_{\bm{\zeta}}^{\top},
    \end{align}
    where   $S= vec \left[ \mathcal{B} (P_{\Theta} \odot (\mu_{\mathbf{v}} \mu_{\mathbf{v}}^\top) ) \mathcal{B}^\top\right]$, with $\mu_{\bf{v}} = - m_0 {\mathbf{C_v}}(I -\mathbf{A})^{-2}\mathbf{B}$.
\end{enumerate}
\end{corollary}
\begin{pf} 
For the first statement, we recall that $\mathbf{M}_{11}(z)$ is the transfer function from 
$y_0$ to $\mu_{\bm{\zeta}}$.
Since the input $y_0$ converges to a ramp signal with slope $m_0$, we can use the final value theorem to conclude that, if $\mathbf{M}_{11}(z)$ has two or more zeros at $z=1$, then
$$\mu_{\bm{\zeta}} =\lim_{z\rightarrow 1} (z-1) \mathbf{M}_{11}(z)\frac{m_0 z}{(z-1)^2} =0.$$
On the other hand, if $\mathbf{M}_{11}(z)$ has only one zero at $z=1$, then
$$\mu_{\bm{\zeta}} =\lim_{z\rightarrow 1} (z-1) \mathbf{M}_{11}(z)\frac{m_0 z}{(z-1)^2} =m_0 \left. \frac{\mathbf{M}_{11}(z)}{z-1} \right|_{z=1}.$$
We can write
\begin{align*}
    \frac{\mathbf{M}_{11}(z)}{z-1} =&
{\mathbf{D}_{\bm{\zeta}}} z^{-1}+({\mathbf{D}_{\bm{\zeta}}}+{\mathbf{C}_{\bm{\zeta}}}\mathbf{B})z^{-2}\\
+&({\mathbf{D}_{\bm{\zeta}}}+{\mathbf{C}_{\bm{\zeta}}}\mathbf{B}+{\mathbf{C}_{\bm{\zeta}}}\mathbf{A}\mathbf{B})z^{-3}\\ +&({\mathbf{D}_{\bm{\zeta}}}+{\mathbf{C}_{\bm{\zeta}}}\mathbf{B}+{\mathbf{C}_{\bm{\zeta}}}\mathbf{A}\mathbf{B}+{\mathbf{C}_{\bm{\zeta}}}\mathbf{A}^2\mathbf{B})z^{-4}+\cdots
\end{align*}
Evaluating at $z=1$, and reordering terms we can write
\begin{align*}
    &\left.\frac{\mathbf{M}_{11}(z)}{z-1} \right|_{z=1} =  \\
   &\lim_{k \rightarrow \infty} \left[
k({\mathbf{D}_{\bm{\zeta}}}+\sum_{i=1}^{k-1}{\mathbf{C}_{\bm{\zeta}}}\mathbf{A}^{i-1}\mathbf{B})-\sum_{i=1}^{k-1}i \; {\mathbf{C}_{\bm{\zeta}}}\mathbf{A}^{i-1}\mathbf{B} \right].
\end{align*}
Note that
${\mathbf{D}_{\bm{\zeta}}}+\sum_{i=1}^{k-1}{\mathbf{C}_{\bm{\zeta}}}\mathbf{A}^{i-1}\mathbf{B}$ converges exponentially fast to ${\mathbf{C}_{\bm{\zeta}}} (I-\mathbf{A})^{-1} \mathbf{B} + {\mathbf{D}_{\bm{\zeta}}}=0$ when $k\rightarrow \infty$. Hence, we have that 
\begin{equation}
\mu_{\bm{\zeta}} =   -{\mathbf{C}_{\bm{\zeta}}} \sum_{i=1}^{\infty} i \; \mathbf{A}^{i-1}\mathbf{B} m_0 = - m_0 {\mathbf{C}_{\bm{\zeta}}}(I -\mathbf{A})^{-2}\mathbf{B}.       
\end{equation}
For the second statement, we recall that $\mathbf{M}_{21}(z)$ is the transfer function from 
$y_0$ to $\mu_{\mathbf{v}}$. Thus, an analogous analysis to the one above allows to conclude that, if $\mathbf{M}_{21}(z)$ has two or more zeros at $z=1$, then  $\mu_{\mathbf{v}}=0$, and if $\mathbf{M}_{21}(z)$ has only one zero at $z=1$, then 
$\mu_{\mathbf{v}} =   - m_0 {\mathbf{C_v}}(I -\mathbf{A})^{-2}\mathbf{B}.
$
Also, since the platoon is MSS, $S(k)$ in \eqref{ec:mss_varianza_estado_lg} converges to
$    S= vec \left[ \mathcal{B} (P_{\Theta} \odot (\mu_{\mathbf{v}} \mu_{\mathbf{v}}^\top) ) \mathcal{B}^\top\right],$ which is zero when $\mu_{\mathbf{v}} =0$ and thus, 
 the recursion in \eqref{ec:mss_varianza_estado_lg} converges to zero since   $\rho((\mathbf{A} \otimes \mathbf{A})+\Delta)<1$ holds. Given that $\mathbf{X}=vec \lbrace P_{\mathbf{X}}\rbrace$, and $P_{\bm{\zeta}}= \mathbf{C}_{\bm{\zeta}} P_{\mathbf{X}} \mathbf{C}_{\bm{\zeta}}^{\top}$ we conclude that $P_{\bm{\zeta}}=0$ in that case. On the other hand, if $\mu_{\mathbf{v}} \neq 0$, then 
 from \eqref{ec:mss_varianza_estado_lg} we have that $\mathbf{X}(k)$ converges to $\mathbf{X}$, given by  
$\mathbf{X}=\left[ I - (\mathbf{A} \otimes \mathbf{A}) -\Delta\right]^{-1}S$, with $S$ as above.
From \eqref{ec:varianza_lg}, and given that $\mathbf{X}=vec \lbrace P_{\mathbf{X}}\rbrace$, we obtain \eqref{ec:var_estacionaria}.
\qed
\end{pf}

Corollary \ref{cor:conv_to_zero}
 states conditions for which the statistics of the error converge or not to zero. We also explicitly provide such stationary values for both, mean and covariance matrix. 
 It is clear then that the chosen strategy to deal with data loss plays an important role since it could affect the MSS and the convergence properties of the platoon statistics. Naturally, it is preferable to choose a strategy that ensures zero stationary statistics for the error. This was observed via simulation results in \cite{gordon2021comparison}.      
Corollary \ref{cor:conv_to_zero} provides an analytical explanation for such phenomena and the specific limiting values.

\subsection{Mutually independent channels}
In this section, we consider the case where the communication channels are assumed to be independent and derive simplified conditions for MSS. Hence, we consider the following assumption:
\begin{assumption}
\label{ass:ind}
We assume that the random processes $\theta_i$, for $i=1,2,\dots N$, are mutually independent. Thus, 
the covariance matrix of $\Theta(k)$ is given by  $P_\Theta = diag(p_1(1-p_1), \dots, p_N(1-p_N)). $
\end{assumption}


\subsubsection{Heterogeneous platoon analysis}
To present our result we define $\alpha_i=A_i+p_iB_iC_{v_i}$, and the transfer functions 
$M_{a_i}(z)=C_{\zeta_i} (zI-\alpha_i)^{-1} B_iD_{v_i} p_i + {D_{\zeta_i}}$ and $M_{b_i}(z)=C_{v_i} (zI-\alpha_i)^{-1} B_iD_{v_i} p_i + {D_{v_i}}$.

\begin{theorem} \label{teo:ind_mss_first_agent}
Consider the platoon described in Section \ref{sec:setup}, with the representation given in Section \ref{sec:SS_representation} and consider Assumption \ref{ass:ind}.
Then 
\begin{enumerate}
    \item 
$\mu_{\bm{\zeta}}(k)$ converges if and only if, $\forall i \in \lbrace 1 \dots, N \rbrace \;$,
\begin{equation}\label{eq:cond_mean_ind}
M_{a_i}(1)=0, \;  \text{ and } \;   \max_{i}\rho(\alpha_i)<1, 
\end{equation}
\item  $P_{\bm{\zeta}}(k)$ converges if and only if $\max_{i}\rho(\alpha_i)<1$ and, $\forall i \in \lbrace 1 \dots, N \rbrace \;$,
\begin{equation}\label{eq:cond_var_ind}
M_{b_i}(1)=0, \; \text{ and } \; \max_{i}\rho(\alpha_i \otimes \alpha_i+\delta_i )<1,   
\end{equation}
where $\delta_i=p_i(1-p_i)(B_i \otimes B_i)(C_{v_i} \otimes C_{v_i})$.
\end{enumerate}
Consequently,  the platoon is mean square stable if and only if  \eqref{eq:cond_mean_ind} and \eqref{eq:cond_var_ind} are met.
 \end{theorem}
 \begin{pf}
 See appendix \ref{pr:teo:ind_mss_first_agent}
\end{pf}

\subsubsection{Homogeneous platoon analysis}
Here we consider a homogeneous case, that is, a platoon where every agent has the same dynamics. 
In a lossy communication setup, homogeneity is not only considered in the  dynamics of the vehicles but also in the communication channels' statistics. 
Hence, here we consider the following assumption:
\begin{assumption}
\label{ass:homo}
The system described in \eqref{ec:estado_lg}, \eqref{ec:error_lg} and \eqref{ec:entrada_canal_lg} holds with $A_i=A$, $B_{i}=B$, $C_{\zeta_i}=C_{\zeta}$, $C_{y_i}=C_{y}$, $C_{v_i}=C_{v}$, $D_{v_i}=D_{v}$ and $D_{\zeta_i}=D_{\zeta}$, that is, every vehicle is described by the same constant matrices ($F_i=F$). Additionally, every inter-vehicle communication channel has the same probability of successful transmission, that is, we set $p_i=p$, where $0 < p \leq 1$.
\end{assumption}
For the homogeneous setup, we define $\alpha=A+pBC_{v}$ and the transfer functions
$M_{a}(z)=C_{\zeta} (zI-\alpha)^{-1} BD_v p + {D_{\zeta}}$ and $M_{b}(z)=C_{v} (zI-\alpha)^{-1} BD_v p + {D_{v}}$.

\begin{corollary}\label{cor:ho_mss_first_agent}
Consider the platoon described in Section \ref{sec:setup}, with the representation given in Section \ref{sec:SS_representation}, and considering Assumption \ref{ass:ind} and \ref{ass:homo}.
Then 
\begin{enumerate}
    \item  
$\mu_{\bm{\zeta}}(k)$ converges if and only if
\begin{equation}\label{eq:cond_mean_ho}
M_{a}(1)=0 \; \text{ and } \;   \rho(\alpha)<1, 
\end{equation}
\item  $P_{\bm{\zeta}}(k)$ converges if and only if $\rho(\alpha)<1$ and
\begin{equation}\label{eq:cond_var_ho}
M_{b}(1)=0 \; \text{ and } \; \rho(\alpha \otimes \alpha+\delta )<1,   
\end{equation}
where $\delta=p(1-p)(B \otimes B)(C_{v} \otimes C_{v})$.
\end{enumerate}
Consequently,  the platoon is mean square stable if and only if  \eqref{eq:cond_mean_ho} and \eqref{eq:cond_var_ho} are met.

\end{corollary} 

\begin{pf}
Straightforward from Theorem \ref{teo:ind_mss_first_agent}.
\end{pf}

Conditions in Theorem \ref{teo:ind_mss_first_agent} and Corollary \ref{cor:ho_mss_first_agent} are  similar in form 
to those in Theorem \ref{theo:MSS}, but applied to simpler matrices.
The results in Corollary \ref{cor:conv_to_zero} are also applicable for the setup with independent channels. It is not difficult to see that the transfer functions required to have two zeros at $z=1$ in order to ensure convergence to zero for the statistics of the tracking error are  $M_{a_i}(z)$ and $M_{b_i}(z)$ for the heterogeneous case and $M_{a}(z)$ and $M_{b}(z)$ for the homogeneous one.

\section{Numerical Examples}
\label{sec:simulation}
In this section, we present numerical tests of the MSS conditions derived in Section \ref{sec:MSS_conditions}. 
Through simulations, different platoon behaviors are presented and discussed. 
Also, we briefly discuss the relationship between MSS and string stability. 
Here we use admissible controllers that stabilize each closed-loop system (with perfect communication). However, in view of the posterior string stability discussion, we also require the platoon with ideal communication to be string stable in the deterministic case, which requires that $||T_i(z)||_{\infty}\leq 1$ (see e.g. \cite{liang2000string}).

The platoon behaviors are presented in figures that use a color code. The first follower is marked in dark blue, while the last follower is marked in dark red. The vehicles in between follow the pattern shown in the color bar at the top of each figure. Each color (from left to right) is mapped with the corresponding increasing index of the vehicles along the string.

\begin{figure}[!t]
\centering
  \begin{tabular}{c}
    \includegraphics[width=.65\columnwidth]{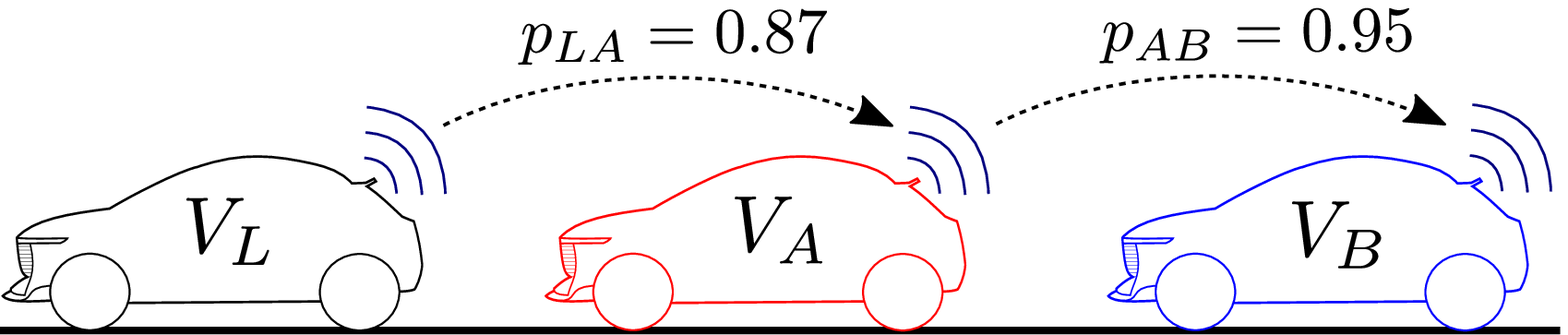}   \\
    \includegraphics[width=.95\columnwidth]{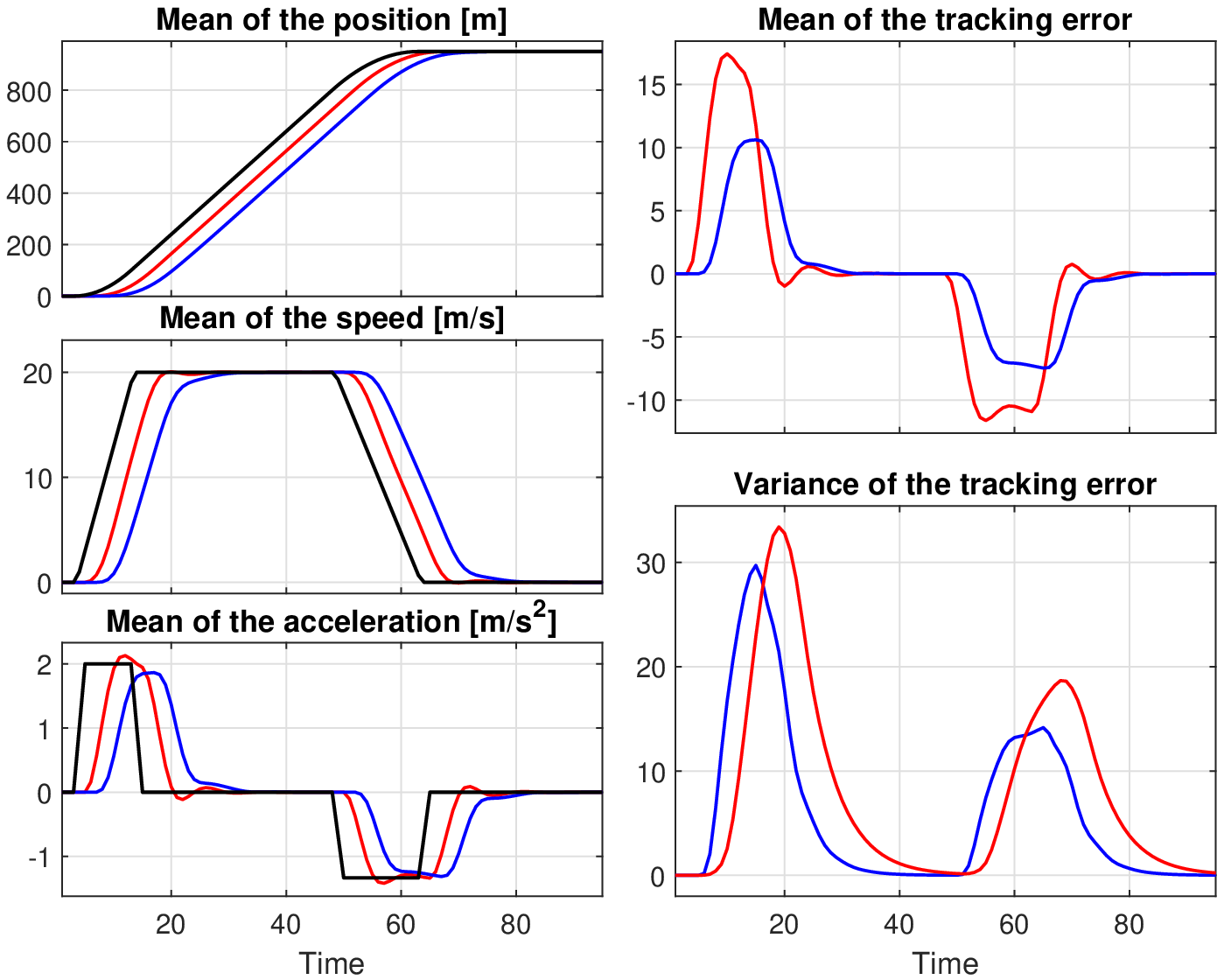} 
  \end{tabular}
  \caption{Behavior of a platoon composed of: leader (black), vehicle $V_A$ (red), and vehicle $V_B$ (blue).}
  \label{plot:LAB_platoon}
\end{figure}

\subsection{MSS analysis with two followers} \label{subs:two_followers}

\begin{figure}[!t]
\centering
  \begin{tabular}{c}
    \includegraphics[width=.65\columnwidth]{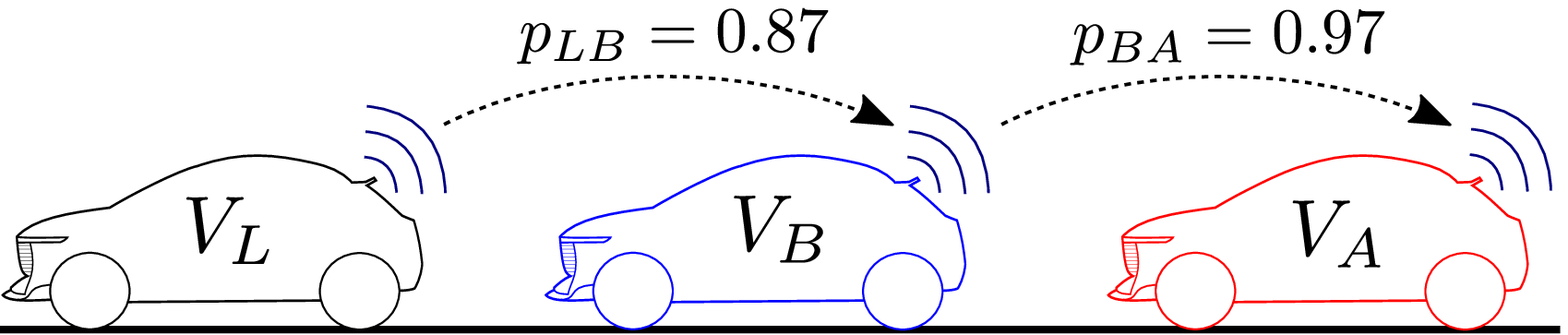}   \\
    \includegraphics[width=.95\columnwidth]{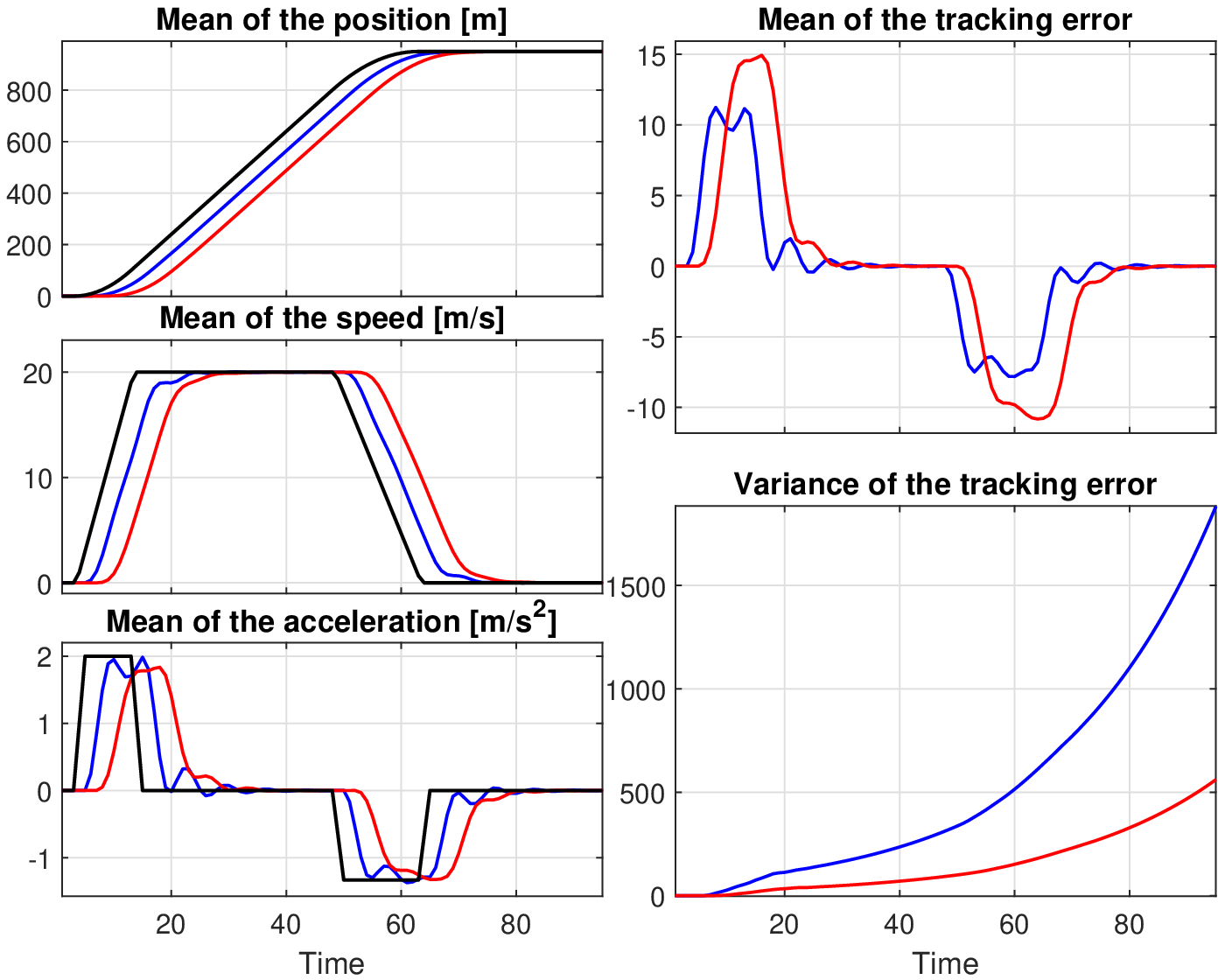} 
  \end{tabular}
  \caption{Behavior of a platoon composed of: leader (black), vehicle $V_B$ (blue), and vehicle $V_A$ (red).}
  \label{plot:LBA_platoon}
\end{figure}

\textit{Influence of the vehicles location}:
Here we consider a heterogeneous platoon with two followers (labeled as $V_A$ and $V_B$) and a leading vehicle (labeled as $V_L$). Each vehicle has the following plant and controller
\begin{align*}
    G_{V_A}(z)&=\frac{1}{z-1},\quad K_{V_A}(z)=\frac{z}{(z-1)(z-0.7)} \\
    G_{V_B}(z)&=\frac{1.2}{z-1},\quad K_{V_B}(z)=\frac{1.33 z }{(z-1)(z-0.88)}.
\end{align*}
For simplicity, we assume that the arrival rates only depend on the transmitting devices and thus we set $p_1=0.87$. In both vehicles, the strategy used is as in Fig. \ref{fig:strat_example}.c) and the time headway is set to $h=3.8$. The lead vehicle starts from rest and then it applies an acceleration of 2 [$m/s^2$] till it reaches a constant speed of 20 [$m/s$]. Finally, from instant $k=50$ the leader brakes to a stop with a deceleration of $-1.33$ [$m/s^2$]. In Fig. \ref{plot:LAB_platoon}, vehicle $V_A$ is in position $i=1$ and vehicle $V_B$ is in position $i=2$. With this setup, the mean and variance of the tracking error converge and the means of the position, speed and acceleration exhibit a good performance. On the other hand, if the followers switch positions, the variance of the tracking error does not converge in the mean square sense. This behavior is shown in Fig. \ref{plot:LBA_platoon}. 
The MSS of the platoon is affected by the location of the vehicles in the string, this is because the inter-vehicle communication between two vehicles depends on the transmitting and receiving devices of such vehicles. Changing the vehicles' location changes, in general, the communication features.

\begin{figure}[t!]
    \begin{center}
    \includegraphics[width=0.99\columnwidth]{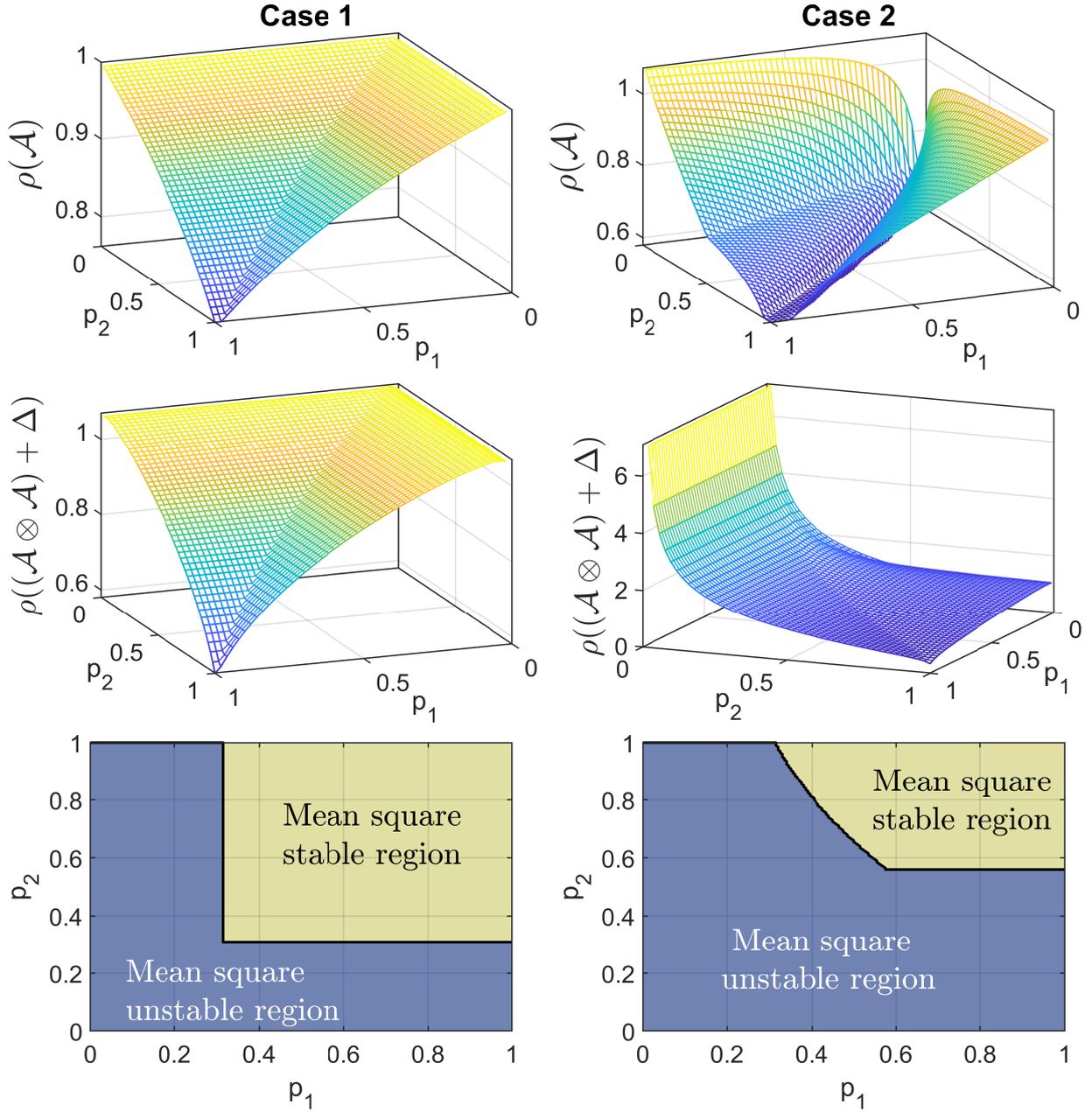}
    \caption{MSS surface for: case 1 (left column) - independent case, case 2 (right column) - dependent case.}
    \label{plot:surf_plots}
    \end{center}  
\end{figure}

\textit{Influence of the arrival rates:}
We consider a platoon with
the strategy in Fig. \ref{fig:strat_example}.b) and time headway  $h=3.8$.
Here we analyze the effect on MSS of the probabilities $p_i$ for two cases. The first one considers spatially independent channels and systems $F_i$ not depending on other channel statistics. The plant and controller are those for Vehicle A above. The second case considers spatially correlated channels and systems $F_i$ depending on other channel statistics. For this case we set $G_1(z)=G_2(z)=G_{V_A}(z)$, $K_1(z)=K_{V_A}(z)/p_1$ and $K_2(z)=2K_{V_A}(z)/(p_1+p_2)$.
In the first column of Fig. \ref{plot:surf_plots} (case 1) we present the results for the independent channels case. Two surfaces reveal the influence between the arrival rates $p_1$ and $p_2$ on the spectral ratio conditions for MSS for the mean and the variance. The spectral radius is larger for low values of $p_1$ and $p_2$, and decreases as long as the probabilities approach to 1, which is expected. In the third graphic, a region in the plane $p_1$ vs $p_2$ where MSS is achieved is presented. It can be seen that there is no interplay between $p_1$ and $p_2$ in this case. We can also notice that there are minimum values for $p_1$ and $p_2$ below which MSS is not possible.
The second case results are given in the second column of Fig. \ref{plot:surf_plots}. The resulting surfaces reveal an important interplay between $p_1$ and $p_2$. Again, higher probabilities reduce the spectral radius value. The MSS region now clearly exhibits an interaction between channels statistics to achieve MSS.

\subsection{MSS analysis of homogeneous platoons} \label{subs:homogeneneous}
\begin{figure}[t!]
    \begin{center}
    \includegraphics[width=0.99\columnwidth]{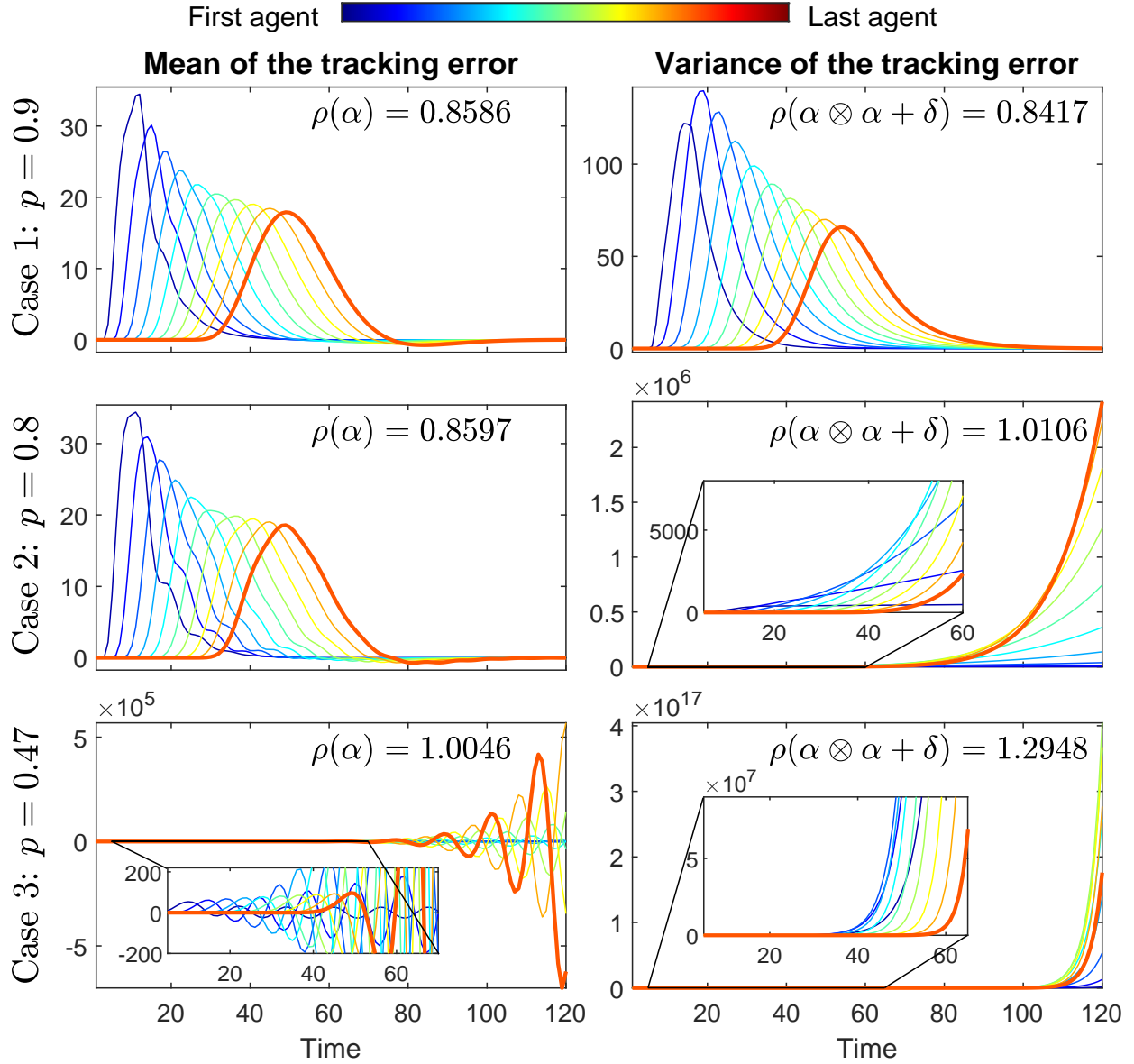}
    \caption{MSS analysis of homogeneous platoons. Plots in the $i$-th row refer to the results for case $i$.}
    \label{plot:homogeneous_sim}
    \end{center}  
\end{figure}

Consider a homogeneous platoon of $N=10$ followers. We set the time-headway $h=4$, $G(z)=(z-1)^{-1}$ and
$$K(z)=\frac{0.27 z (z+0.88)}{(z-1)(z-0.79)(z-0.8)}.$$
The compensation strategy implemented is the one in Fig. \ref{fig:strat_example}$(c)$. In Fig. \ref{plot:homogeneous_sim} the leader reference is a ramp (vehicle moving with constant speed of $35 \left[ m/s \right]$). With this setup, for every $0<p\leq 1$, conditions $M_a(1)=0$ and $M_b(1)=0$ are achieved (see Corollary \ref{cor:ho_mss_first_agent}). For instance, with $p=0.9$ we have
\begin{align*}
    {M}_{a}(z)&= \frac{(z-1)^2(z+0.79)(z-0.1)}{(z+0.39)(z-0.85)(z^2-0.84z+0.56)} \notag \\
    {M}_{b}(z)&=\begin{bmatrix}
       \frac{(z-1)^3 (z+0.79)}{(z+0.39) (z-0.86) (z^2 - 0.84z + 0.56)} \\
       \frac{0.24 z (z-1)^2 (z-0.88)}{(z+0.39) (z-0.8) (z-0.86) (z^2 - 0.84z + 0.56)}
    \end{bmatrix}.
\end{align*}
It can be observed that both ${M}_{a}(z)$ and ${M}_{b}(z)$ have at least two zeros at $z=1$.
\\
By varying the arrival rate we have different results. In Fig. \ref{plot:homogeneous_sim} we show three different cases.
When the probability of success is set to $p=0.9$, the mean and variance of the tracking error converge uniformly as time approaches infinity. The spectral radius conditions are met with the values $\rho(\alpha)=0.8586$ and $\rho(\alpha \otimes \alpha+\delta)=0.8417$. Also, the system achieves zero error in steady-state, which is also verified through Corollary \ref{cor:conv_to_zero}. 
 When the probability of success is diminished to $p=0.8$, the platoon becomes mean square unstable. In this example $\rho(\alpha)=0.8597$ and thus, the convergence of the mean is ensured, but the variance does not converge since $\rho(\alpha \otimes \alpha+\delta)=1.0106 \nless 1$. Thus, the instability is due to the variance behavior. 
Since the platoon is homogeneous, the unstable behavior of the variance can be observed from the first follower. 
For the third case, neither the mean nor the variance converge. This behavior is forced by choosing a poor channel with a probability of success $p=0.47$. The spectral radius conditions are greater than one ($\rho(\alpha)=1.0046$, $\rho(\alpha \otimes \alpha+\delta)=1.2948$).

\subsection{MSS analysis of a heterogeneous platoon}
\label{subs:heterogeneous}
\begin{figure}[t!]
    \begin{center}
    \includegraphics[width=0.99\columnwidth]{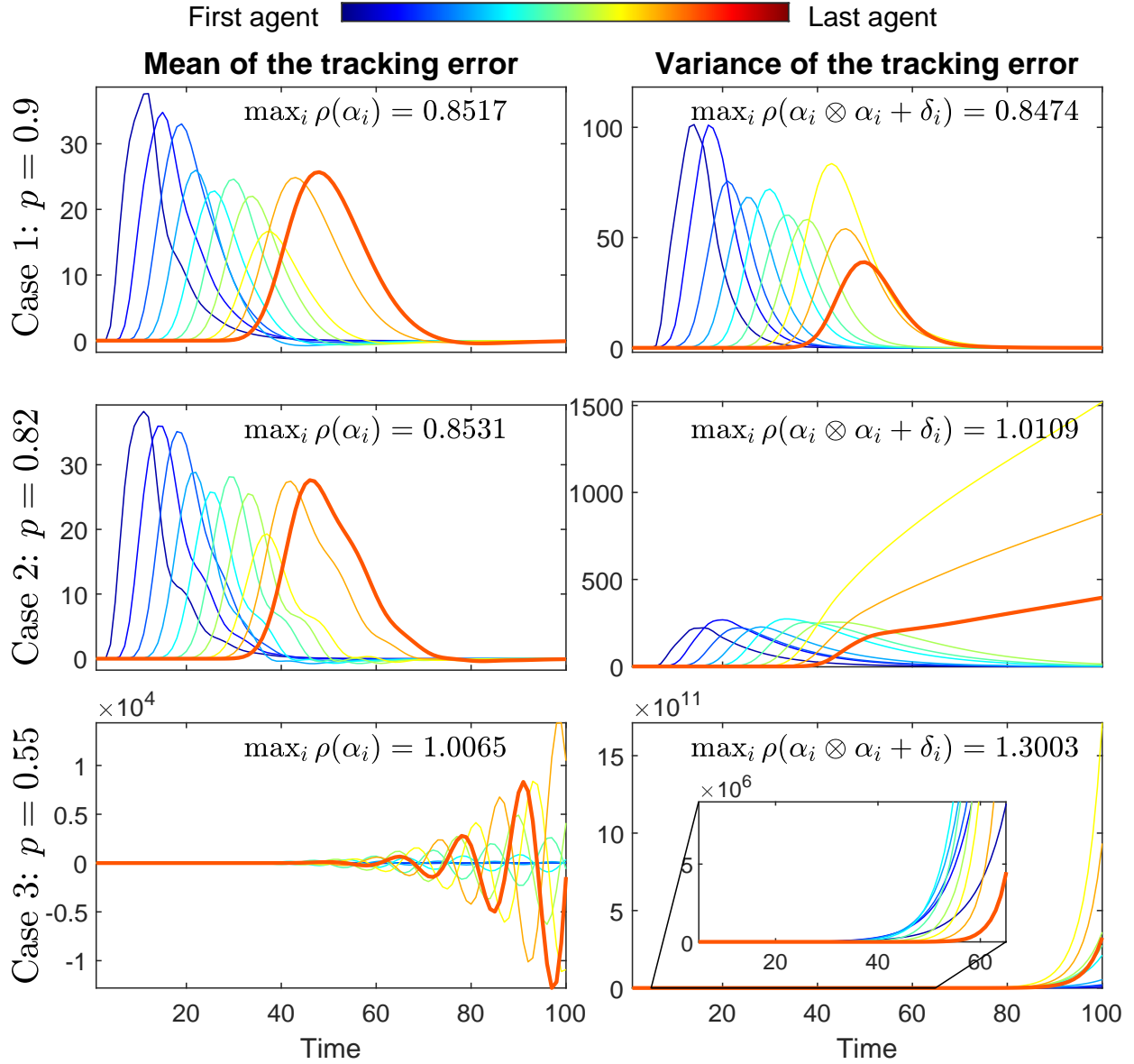}
    \caption{MSS analysis of heterogeneous platoons. Plots in the $i$-th row refer to the results for case $i$.}
    \label{plot:heterogeneous_sim}
    \end{center}  
\end{figure}

Consider a heterogeneous platoon of $N=10$ followers with independent channels, where each vehicle has its own dynamic and controller, but share the same structure. 
The model of the vehicles is characterized by  $ G_i(z)=g_i/(z-1)$, where $g_i$ is the gain of the plant. The controllers structure is as follow
\begin{equation}
    K_i(z) =\frac{(k_i/(1+h))z\ (z+c_i)}{(z-1)(z+\lambda_i)(z-(h/(1+h)))}
\end{equation}
where $k_i$, $c_i$ and $\lambda_i$ are the gain, zero and pole of the controller, respectively. 
The leader reference is a ramp (moving with constant speed of $35 \left[ m/s \right]$). All agents use the compensation strategy in Fig. \ref{fig:strat_example}$(c)$, time-headway with $h=4$, and the channels have the same arrival rate $p$. In Fig. \ref{plot:heterogeneous_sim} we present three different cases of convergence for different values of $p$.
 For $p=0.9$, the statistics of the tracking error converge to zero in the mean square sense. Here we have $ \max_{i}\rho(\alpha_i)= 0.8517$ and $\max_{i}\rho(\alpha_i \otimes \alpha_i+\delta_i)=0.8474$,
respectively. In contrast with the homogeneous platoon, the statistics of each follower have different transient and peak responses. 
By decreasing the probability of success to $p=0.82$, the vehicle in the position $i=8$ becomes mean square unstable. The instability occurs only in the variance of the tracking error. The conditions obtained from the mean square analysis are
$ \max_{i}\rho(\alpha_i)= 0.8531$ and  $\max_{i}\rho(\alpha_i \otimes \alpha_i+\delta_i)=1.0109$. Notice that the non-convergence of the $8-th$ vehicle forces the non-convergence of the upcoming vehicles. Hence, the whole platoon becomes mean square unstable.
For $p=0.55$, both conditions
$\max_{i}\rho(\alpha_i)= 1.0065$ and $\max_{i}\rho(\alpha_i \otimes \alpha_i+\delta_i)=1.3003$ 
exceed the value of one. Therefore, the mean and variance of the tracking error dramatically increase in amplitude.

A platoon with non-constant time headway or whose communication channels have different probabilities of success is also a heterogeneous platoon. The expected behavior is similar to the one presented above.

\subsection{Analysis of strategies for data-loss compensation}
Now we will illustrate the fact that even if the ideal communication platoon is designed to reach zero steady-state error according to Assumption \ref{ass:transfer_function}, when data-loss occurs, the influence of the chosen strategy could change the convergence properties. We consider the controller and the plant of the homogeneous case in Section \ref{subs:homogeneneous}.

\begin{figure}[!t]
    \begin{center}
    \includegraphics[width=0.95\columnwidth]{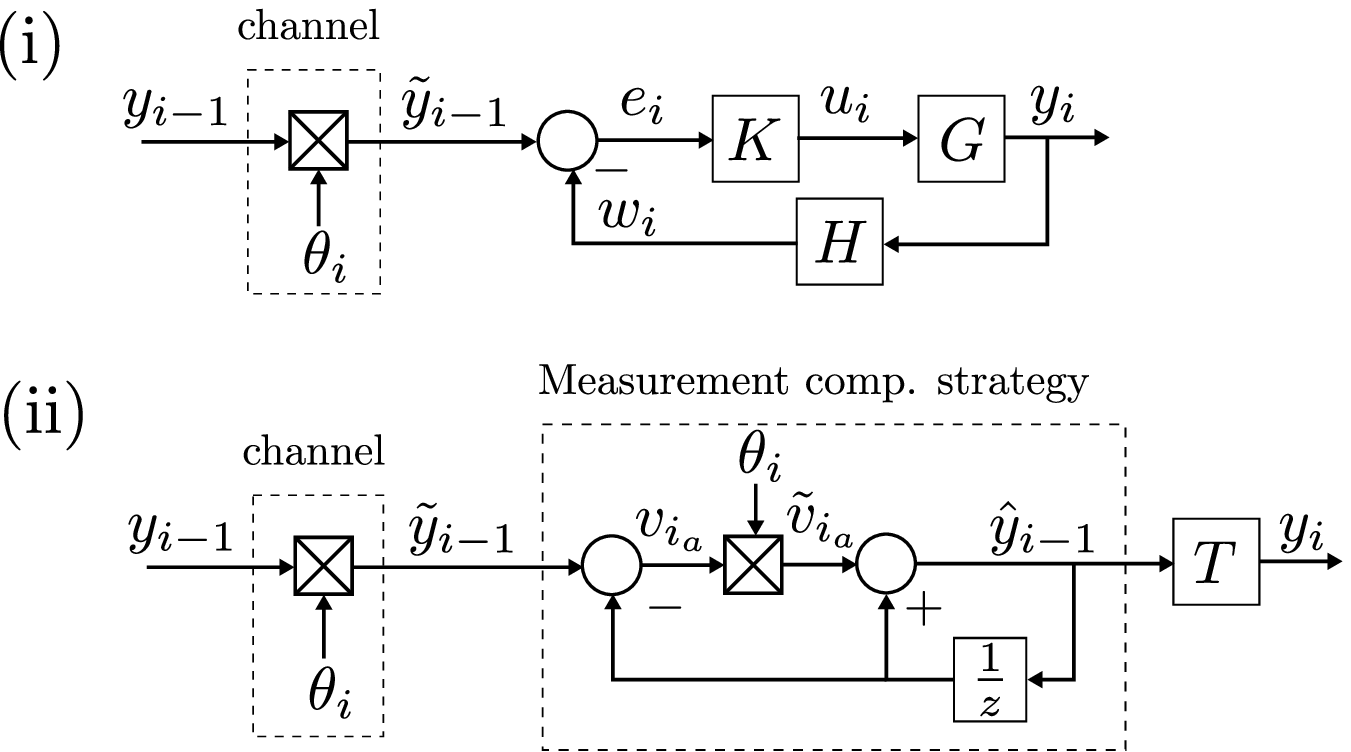}
    \caption{Example of compensation strategies with non-zero steady-state statistics.}
    \label{fig:strat_special}                     
    \end{center}                             
\end{figure}

\begin{figure}[!t]
    \begin{center}
    \includegraphics[width=0.99\columnwidth]{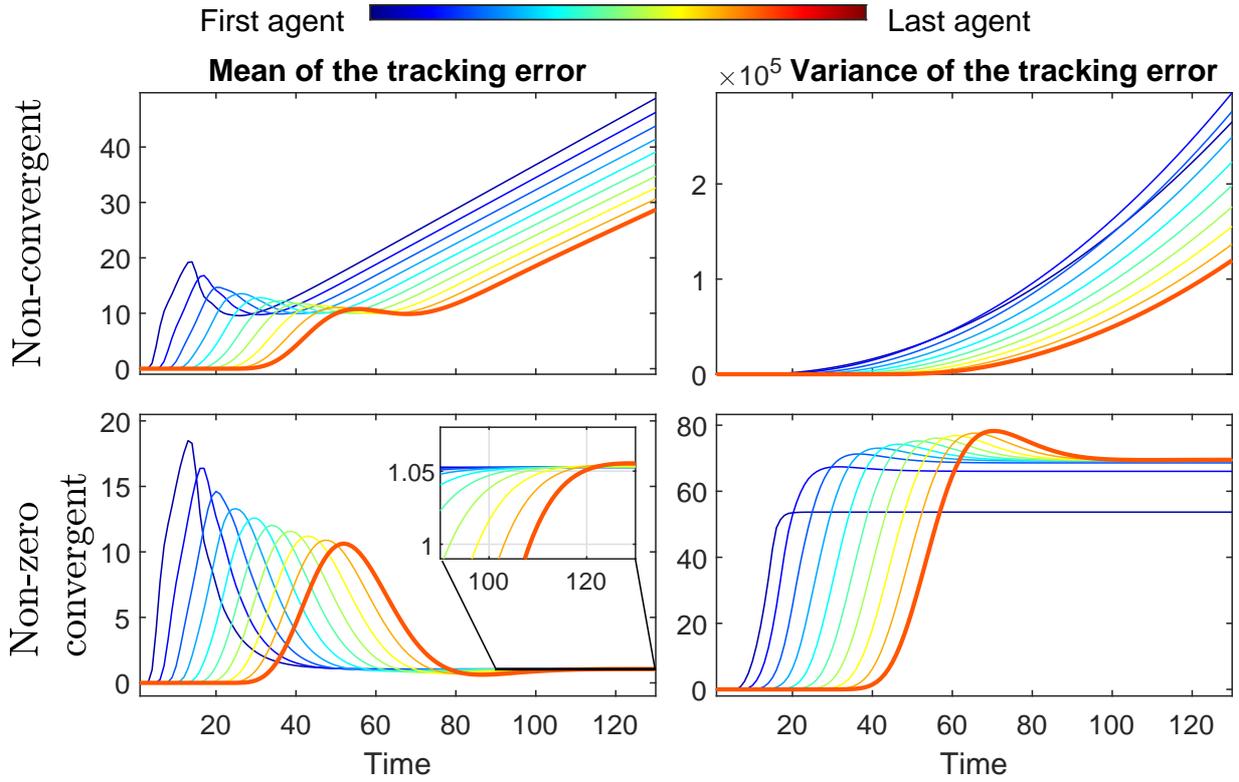}
    \caption{First row: behavior of a platoon with non-convergent strategies. Second row: behavior of a platoon with non-zero convergent strategies.}
    \label{plot:special_strat}
    \end{center}                             
\end{figure}

\textit{Example of a non-convergent strategy:} 
Consider a homogeneous platoon with a data-loss compensation strategy in which the missing data is replaced by zero (see first row of Fig. \ref{fig:strat_special}). In Fig. \ref{plot:special_strat} we show the simulation results for $N=10$ vehicles, $p=0.98$ and $h=4$. The transfer function ${M}_{a}(z)$ obtained is
\begin{equation*}
    {M}_{a}(z)= \frac{(z+0.78)(z^2-1.99z+0.988)}{(z-0.853)(z^2-0.356z+0.446)}
\end{equation*}
and $M_b(z)=1$. Even if the channel is almost ideal, and the spectral radius conditions are compatible with MSS,  the system cannot track the reference because of the missing zeros at $z=1$, as predicted by Corollary \ref{cor:conv_to_zero}, and MSS is not achieved.

\textit{Example of a non-zero convergent strategy:}    
Consider a platoon with a to-hold compensation strategy in the measurement signal as depicted in the second row of Fig. \ref{fig:strat_special}. In Fig. \ref{plot:special_strat} we show the numerical results of a homogeneous platoon with $p=0.95$ and $h=4$. The conditions obtained are $\rho(\alpha)=0.8535$ and $\rho(\alpha \otimes \alpha+\delta)=0.7284$. The transfer functions obtained are
$$
    {M}_{a}(z) = \frac{(z-1)(z-0.995)(z+0.71)(z+0.027)}{(z-0.05)(z-0.853)(z^2-0.356z+0.446)}
$$ 
and ${M}_{b}(z) = (z-1)(z-0.05)^{-1}$. As seen, the spectral radius conditions are satisfied, however, $M_{a}$ and $M_b$ have only one zero at $z=1$. Consequently, the statistics converge to a value different from zero.

\textit{Example of a zero convergent strategy:}    
Any of the compensation strategies employed in section \ref{subs:two_followers}, \ref{subs:homogeneneous} and \ref{subs:heterogeneous} are zero convergent.

\subsection{String Stability discussion} \label{sec:MSSvsSS}
\begin{figure}[!t]
    \begin{center}
    \includegraphics[width=0.99\columnwidth]{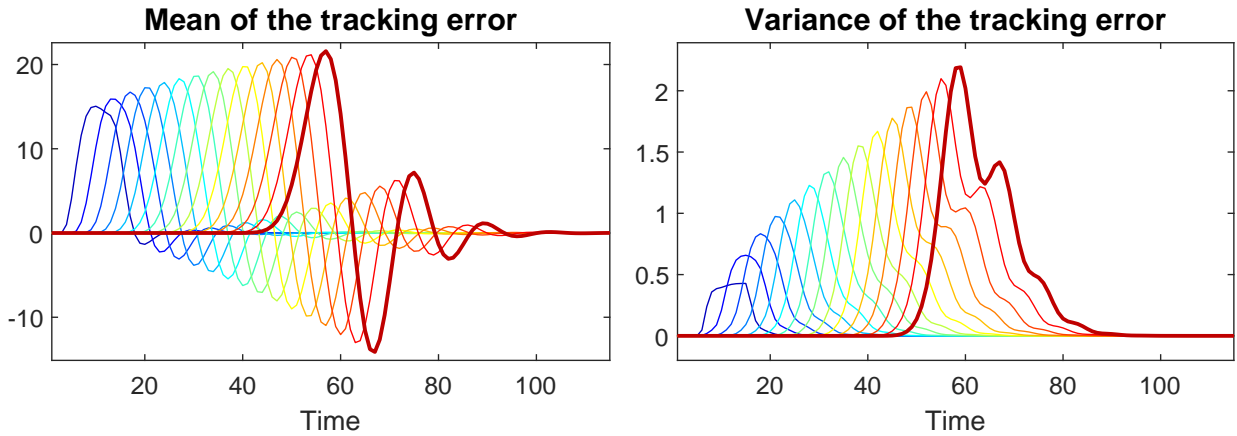}
    \caption{String unstable behavior}
    \label{plot:ho_string}
    \end{center}                             
\end{figure}

String stability is a property that guarantees that disturbances are not amplified along the string of vehicles as they propagate. String stability  is a stronger property when compared with internal stability, since, in addition to the convergence in time, it demands the signals of interest to remain uniformly bounded when the number of agents increases. 
Although in a deterministic setting string stability has been extensively studied, results are scarce for stochastic scenarios \cite{xu2018mean,li2019string,feng2019string,li2021event}. Nevertheless, some works suggest  that a platoon with stochastic signals involved is string stable if the mean and the variance of the error exhibit a string stable behavior \cite{vegamoor2021string,elahi2021h,zhao2020stability}.
For instance, the first row in Fig. \ref{plot:homogeneous_sim} where a homogeneous platoon is considered, shows a string stable behavior for both the mean and the variance. Conversely, if we use the parameters $N=15$, $p=0.95$ and $h=3$ we have a string unstable behavior, as shown in Fig. \ref{plot:ho_string}. Note that both the mean and variance in Fig. \ref{plot:ho_string} are mean square stable and the statistics converge to zero, however, the string unstable behavior can be appreciated in the peak magnitude of the statistics. As more vehicles are added to the string, said amplitudes reach higher peak values. Of course, this behavior is undesirable in platooning. 

Some interesting results about platooning with random data loss can be found in \cite{acciani2021stochastic}, where the authors first impose a string stability behavior for the mean and then they minimize the corresponding variance. However, this two-step approach should be applied judiciously because there are cases in which depending on the platooning setup, the mean behaves as string stable, but the variance is not even mean square stable, such as in the second row of Fig. \ref{plot:homogeneous_sim}. 
We believe that, for a proper approach to study string stabilization in an stochastic setting, both the mean and variance of the tracking error must be required to behave in string stable fashion. Hence, the necessary and sufficient conditions for MSS derived in this work would be necessary conditions for string stabilization of platooning over lossy channels. 

\section{Conclusions and future work}
In this article, we studied the mean square stabilization of platoons of vehicles with lossy V2V communication channels. When the dropout occurs, a compensation strategy was used to replace the missing data or modify an internal signal of the closed-loop. 
We derived necessary and sufficient conditions for MSS considering heterogeneous platoons and spatially correlated communication channels, guaranteeing the convergence of the statistics of the tracking errors. We found that the channels' qualities and the employed compensation strategy play a crucial role for MSS. Additionally, we derived a condition that guarantees that said convergence reaches zero in steady-state. We also present simplified conditions for the case when the channels are independent. Through simulation examples, we tested the performance of homogeneous and heterogeneous platoons with different probabilities of success and compensation strategies. 
We also provide a brief discussion about the importance of considering MSS as a basic requirement in order to study string stabilization over lossy channels.

As part of our future work, we will consider extending the MSS analysis to platoons with different communication topologies (e.g. leader-predecessor following) and control schemes (e.g. speed control). Using the MSS conditions to properly design controllers and compensation strategies is also part of the future work. Obtaining conditions for string stabilization over lossy channels is also in the scope of our future work.

\bibliographystyle{plain}
\bibliography{mybib}

\appendix
\section{Appendix}
\subsection{Proof of Lemma \ref{lema_ss}} \label{prf:lema_ss}
By replacing the input position of each agent \eqref{ec:position_single} in \eqref{ec:error_single} and \eqref{ec:canal_single}, it is not difficult to show that the open loop of the concatenated system can be described by the following state-space representation
\begin{equation}
\label{ec:sys_original}
    \begin{bmatrix}
    \mathbf{x}(k+1) \\ \bm{\zeta}(k) \\ \mathbf{v}(k)
    \end{bmatrix} =
    \begin{bmatrix}
    \mathcal{A} && \bf{0} && \mathcal{B} \\
    \mathbf{C}_{\bm{\zeta}} && \mathbf{D}_{\bm{\zeta}} && \bf{0} \\
    \mathbf{C_v} && \mathbf{D_v} && \bf{0} \\
    \end{bmatrix} \ 
    \begin{bmatrix}
    \mathbf{x}(k) \\ y_0(k) \\ \tilde{\mathbf{v}}(k)
    \end{bmatrix},
\end{equation}
where $\mathcal{A}= diag (A_1, \hdots,A_N)$ and $ \tilde{\mathbf{v}}(k)=  \Theta_d(k) \mathbf{v}(k).$
Notice that some matrices of the state-space representation are matrices of zeros due to transfer functions being strictly proper. The cascade structure of the platoon is reflected in the state-space matrices, which are block diagonal ($\mathcal{A}$, $\mathcal{B}$) 
or lower block bidiagonal ($\mathbf{C}_{\bm{\zeta}}$, $\mathbf{C_v}$) matrices.
Due to only the first follower having access to the leader position, matrices $\mathbf{D}_{\bm{\zeta}} $ and $\mathbf{D_v}$ have only one element different from zero. Then, considering that   $\bar{\Theta}_d(k)=\Theta_d(k)-\Upsilon$, $ \breve{\mathbf{v}}(k) = \bar{\Theta}_d(k) \mathbf{v}(k)$, and defining $\mathbf{A} = \mathcal{A}+ \mathcal{B} \Upsilon \mathbf{C_v}$, $\mathbf{B} = \mathcal{B} \Upsilon \mathbf{D_v}$; the lemma is proven straightforwardly from \eqref{ec:sys_original}.
\qed

\subsection{Proof of Lemma \ref{lema_statistics}} \label{pr:lemma_statistics}
The three means in \eqref{ec:media_lg} are obtained directly by applying the expectation operator to \eqref{ec:ss_lg} and considering the fact that the input signal $y_0(k)$ is deterministic. 
The variance of the state is obtained as follows:
\begin{align*}
    &P_{\mathbf{x}}(k+1) \\
    &= \mathcal{E} \big\{ \left[ \mathbf{A}\bar{\mathbf{x}}(k) + \mathcal{B} \bar{\Theta}_d(k)\mathbf{v}(k) \right]
     \left[ \mathbf{A} \bar{\mathbf{x}}(k) + \mathcal{B} \bar{\Theta}_d(k)\mathbf{v}(k) \right]^\top\big\}\\
    &= \mathbf{A}\ \mean{\bar{\mathbf{x}}(k) \bar{\mathbf{x}}(k)^\top} \mathbf{A}^\top + \mathcal{B}\ \mean{\bar{\Theta}_d(k) \mathbf{v}(k) \bar{\mathbf{x}}(k)^\top} \mathbf{A}^\top \\
    &\quad + \mathbf{A}\ \mean{\bar{\mathbf{x}}(k) \mathbf{v}(k)^\top \bar{\Theta}_d(k)^\top} {\mathcal{B}}^\top \\
    &\quad + \mathcal{B}\ \mean{\bar{\Theta}_d(k) \mathbf{v}(k) \mathbf{v}(k)^\top \bar{\Theta}_d(k)^\top} {\mathcal{B}}^\top .
    &&
\end{align*}
Given the assumption of independence (in time) of the channels in $\Theta(k)$, we see that  
\begin{align*}
    &\mean{\bar{\Theta}_d(k) \mathbf{v}(k) \bar{\mathbf{x}}(k)^\top} = \mean{\bar{\Theta}_d(k)} \mean{\mathbf{v}(k) \bar{\mathbf{x}}(k)^\top}  =0,
\end{align*}
since $\mean{\bar{\Theta}_d(k)}=0$. Likewise, $\mean{\bar{\mathbf{x}}(k) \mathbf{v}(k)^\top \bar{\Theta}_d(k)^\top}=0$. 
Using the property in \eqref{ec:prop_schur_def}, we have
\begin{align*}
    &\mean{\bar{\Theta}_d(k) \mathbf{v}(k) \mathbf{v}(k)^\top \bar{\Theta}_d(k)^\top} 
     = P_{\Theta} \odot \mean{\mathbf{v}(k) \mathbf{v}(k)^\top}.
\end{align*}
Since $y_0(k)$ is deterministic, the covariance matrix of the tracking error and the channel input can be written as
\begin{align*}
    P_{\bm{\zeta}}(k) &= \mean{\bar{\bm{\zeta}}(k)\ \bar{\bm{\zeta}}(k)^\top} = \mathbf{C}_{\bm{\zeta}}\ \mean{\bar{\mathbf{x}}(k) \bar{\mathbf{x}}(k)^\top} {\mathbf{C}_{\bm{\zeta}}}^\top\\
    P_{\mathbf{v}}(k) &= \mean{\bar{\mathbf{v}}(k) \bar{\mathbf{v}}(k)^\top} 
    = \mathbf{C_v}\ \mean{\bar{\mathbf{x}}(k) \bar{\mathbf{x}}(k)^T} {\mathbf{C_v}}^\top .  
\end{align*}
Which completes the proof \hfill \qed

\subsection{Proof of Theorem \ref{teo:ind_mss_first_agent}} \label{pr:teo:ind_mss_first_agent}
In this proof, we slightly modify the  notation to explicitly include the number of vehicles of the platoon. This will allow us to write the platoon dynamics recursively in terms of smaller platoons to use mathematical induction and prove our claim. 
Hence, given the representation in Lemma \ref{lema_ss} we write
\begin{align*}
\mathbf{A}  \rightarrow  \mathbf{A}_N &= 
    \left[\begin{array}{c|c}
    \mathbf{A}_{N-1} & \bm{0} \\
    \hline
   \Psi_{N}  & \alpha_N 
    \end{array}\right]\\
\mathbf{B} \rightarrow  \mathbf{B}_N &=
    \begin{bmatrix}
    \beta_1^{\top} & \bm{0} & \cdots & \bm{0} 
    \end{bmatrix}^\top = \left[\begin{array}{c|c}
    \mathbf{B}_{N-1} & \bm{0}
    \end{array}\right]^\top .
\end{align*}
where $\beta_i=p_i B_iD_{v_i}$ and 
$\Psi_{N} = [\bm{0} \; \;  \gamma_N]$ for $N\geq 3$, and $\Psi_{\bf{2}}=\gamma_2$.
We also define, for $N\geq 3$,
\begin{align*}
\varPhi_{\bm{\zeta}_N} &= 
     \begin{bmatrix}
        \bm{0} &&   D_{\zeta_N}C_{y_{N-1}}
    \end{bmatrix},  \textrm { with } \varPhi_{\bm{\zeta}_2}= D_{\zeta_2}C_{y_{1}}, \\
\varPhi_{\mathbf{v}_N} &= 
     \begin{bmatrix}
        \bm{0} &&  D_{v_N} C_{y_{N-1}}
    \end{bmatrix},  \textrm { with } \varPhi_{\mathbf{v}_2} = D_{v_2} C_{y_{1}},
\end{align*}
where $\bf{0}$ is a zero matrix of appropriate dimensions.
We also explicitly include the length of a platoon in the concatenated state, the error and channel input vectors, and write $\mathbf{x_{N}}(k)$, $\bm{\zeta_{N}}(k)$ and $\mathbf{v_{N}}(k)$ respectively. The terms $x_N(k)$ $\zeta_N(k)$ and $v_N(k)$ refer to the state, the error and the channel input of the $N$-th vehicle.  
With this notation, we can write a representation of a platoon with $N$ vehicles in terms of the set of $N-1$ predecessor vehicles, allowing us to analyze the effects of adding a new vehicle to the string. Indeed, it is straightforward to see that
\begin{equation*}
    {\mu_{\mathbf{x_{N}}}}(k) = \left[\begin{array}{c | c}
    {\mu_{\mathbf{x_{N-1}}}}(k)^{\top} \; & \;
    {\mu_{x_{N}}}(k)^{\top}
    \end{array}\right]^{\top},
\end{equation*}
and similarly for the error and channel input means.
Thus, from \eqref{ec:media_estado_lg_ast}, the mean of the alternative state $\mathbf{x_{N}}^*(k)=\mathbf{x_{N}}(k)-\mathbf{x_{N}}(k-1)$ satisfies
${\mu_{\mathbf{x_{N}^* }}}(k+1) = \mathbf{A}_{\bf{N}} {\mu_{\mathbf{x_{N}^*}}} (k) + \mathbf{B}_{\bf{N}} m_0(k).
$
For the first follower we can write ${\mu_{\mathbf{x_{1}^*}}} (k+1)=\alpha_1 {\mu_{\mathbf{x_{1}^* }}}(k) + \beta_1 m_0(k)$
which converges if and only if the condition $\rho(\alpha_1)<1$ holds. Furthermore, we have that  (see proof of Theorem \ref{theo:MSS})
\begin{align}
  \mu_{\bm{\zeta}_1} &= \mu_{\bm{\zeta}_1} + \left(C_{\zeta_1} (I-\alpha_1)^{-1} B_1D_{v_1} p_1 + {D_{\zeta_1}}\right) m_0. \notag \\
    \label{eq:igualdad}
    &= \mu_{\bm{\zeta}_1} + M_{a_1}(1) m_0.
\end{align}
Clearly, the equality in \eqref{eq:igualdad} requires $M_{a_1}(1)=0$ to be valid.
For $N=2$ we can write
\begin{equation}
    {\mu_{x_{2}^*}} (k+1) = \alpha_2 {\mu_{x_{2}^*}} (k) + \Psi_{2} {\mu_{\mathbf{x_{1}}}}^*(k).
\end{equation}
Consequently, ${\mu_{x_{2}^*}} (k)$ converges when $\rho(\alpha_2)<1$ and $\rho(\alpha_1)<1$. Specifically, ${\mu_{x_{2}^*}} (k)$ converges to ${\mu_{x_{2}^*}} =(I-\alpha_2)^{-1}\Psi_2 {\mu_{\mathbf{x_{1}^*}}}$.
From  \eqref{ec:media_error_lg_ast}, the corresponding follower error can be written as $\mu_{{\zeta}_2}(k) = \mu_{{\zeta}_2}(k-1) + C_{\zeta_2}  \mu_{x_2^*}(k) +\varPhi_{\bm{\zeta}_2} \mu_{\mathbf{x}_1^*}(k).$
In the limit $\mu_{{\zeta}_2}:=\lim_{k \rightarrow \infty} \mu_{{\zeta}_2}(k) $ satisfies
\begin{align*}
\mu_{{\zeta}_2}&= \mu_{{\zeta}_2} + C_{\zeta_2}  (I-\alpha_2)^{-1}\Psi_{\bf{2}} {\mu_{\mathbf{x_{1}^*}}} +\varPhi_{\bm{\zeta}_2} \mu_{\mathbf{x}_1^*}\\
&=\mu_{{\zeta}_2} + M_{a_2}(1) C_{y_1}\mu_{{x}_1^*}.
\end{align*}
Since $M_{a_2}(1)=C_{\zeta_2} (I-\alpha_2)^{-1} B_2D_{v_2} p_2 + {D_{\zeta_2}}=0$, the convergence is ensured.
For an arbitrary $N$ we have
\begin{equation}
    {\mu_{x_{N}^*}} (k+1) = \alpha_N {\mu_{x_{N}^*}} (k) + \Psi_{\bf{N}} {\mu_{\mathbf{x^*_{N-1}}}}(k).
\end{equation}
which additionally requires $\rho(\alpha_N)<1$ to converge, yielding $\mu_{{\zeta}_N}= \mu_{{\zeta}_N} + C_{\zeta} \mu_{x^*_N} +\varPhi_{\bm{\zeta}_{N}} \mu_{\mathbf{x}_{N-1}^*}
=\mu_{{\zeta}_N} + M_{a_N}(1) C_{y_{N-1}}\mu_{{x}_{N-1}^*}$,
which requires $M_{a_N}(1)=0$ to ensure the existence of the stationary value $\mu_{{\zeta}_N}$. Clearly $\mu_{{\zeta}_N}$ converges if and only if $\max_i \rho(\alpha_i)<1$ and $M_{a_i}(1)=0$ for every $i=1,2, \dots N$, which proof the first statement in Theorem \ref{teo:ind_mss_first_agent}.

It is easy to prove, mutatis mutandis, that the convergence of $\mu_{\mathbf{v}_N}$ is guaranteed if and only if  $\max_i \rho(\alpha_i)<1$ and $M_{b_i}(1)=0$ for every $i=1,2, \dots N$.

On the other hand, the covariance matrix of the state can be written as
\begin{equation*}
    P_{\bf{x_{N}}}(k) = \left[\begin{array}{c|c}
    P_{\bf{x_{N-1}}}(k) & P_{{\bf{x}}_{N-1}{x}_{N}}(k) \\
    \hline
    P_{{x}_{N}{\bf{x}}_{N-1}}(k)  & P_{x_{N}}(k) 
    \end{array}\right], 
\end{equation*}
Thus, the convergence of the error covariance matrix $P_{\bf{\zeta}_N}(k)=\mathbf{C}_{\bm{\zeta}_N} P_{\bf{x_{N}}}(k) \mathbf{C}_{\bm{\zeta}_N}^{\top},$ requires the convergence of $P_{\bf{x_{N-1}}}(k)$, $P_{x_{N}}(k)$ and $P_{{\bf{x}}_{N-1}{x}_{N}}(k)=P_{{x}_{N}{\bf{x}}_{N-1}}^\top(k)$.
To ease the notation we define
    \begin{align*}
   \Omega_{N}(k)=& B_N \left[ P_{\theta_N} \odot (\mu_{v_{N}}(k)\ {\mu_{v_{N}}(k)}^\top) \right] B_N^\top \\
   R_{N}(k)=&\Psi_{\bf{N}} P_{\mathbf{x_{N-1}}}(k) {\Psi_{\bf{N}}}^{\top} \\
    &+ B_N \left[ P_{\theta_N} \odot (\varPhi_{\mathbf{v}_N} P_{\mathbf{x_{N-1}}}(k)\ {\varPhi_{\mathbf{v}_N}}^{\top}) \right] {B_N}^{\top} \notag \\ 
    \Lambda_N(k) =& \alpha_N P_{x_{N}{\bf{x_{N-1}}}}(k) {\Psi_{\bf{N}}}^{\top} \notag + \Psi_{\bf{N}} P_{{\bf{x_{N-1}}} x_{N}}(k) \alpha_N^{\top} \\
    &+ B_N \left[ P_{\theta_N} \odot (C_{v_N} P_{x_{N}{\bf{x_{N-1}}}}(k)\ {\varPhi_{\mathbf{v}_N}}^{\top}) \right] {B_N}^{\top} \notag \\
    & + B_N \left[ P_{\theta_N} \odot (\varPhi_{\mathbf{v}_N} P_{{\bf{x_{N-1}}} x_{N}}(k)\ {C_{v_N}}^{\top}) \right] {B_N}^{\top} \notag.
\end{align*}
It must be noted that $\Omega_N(k)$, $R_N(k)$ and $\Lambda_N(k)$ converge if and only if $\mu_{v_{N}}(k)$, $P_{\mathbf{x_{N-1}}}(k)$ and $P_{x_{N},{\bf{x_{N-1}}}}(k)$ converge, respectively.

Given the heterogeneous setup with mutually independent channels, we notice that, for $N=1$
\begin{align}
    P_{x_1}(k+1) =& \; \alpha_1 \ P_{x_1}(k)\  \alpha_1^\top + \Omega_1(k)\\
    &+ B_1 \left[ P_{\theta_1} \odot ({C_{v_1}}\ P_{x_1}(k) {{C_{v_1}}}^\top) \right] B_1^\top. \notag 
\end{align}
Using a similar reasoning as in the proof of Theorem \ref{theo:MSS} for the variance convergence, we notice that, given $\rho(\alpha_1)<1$ and $M_{b_1}(1)=0$, $\Omega_1(k)$ converges and thus the matrix $P_{x_1}(k)$ also converges if $\rho((\alpha_1 \otimes \alpha_1) +\delta_1)<1$, where $\delta_1$ is as in Theorem \ref{teo:ind_mss_first_agent}. 

For $N=2$, we have to analyze $P_{x_1}(k)$, $P_{x_2}(k)$ and the cross term $P_{x_1x_2}(k)$. 
Since $P_{\Theta_N}$ is diagonal, then the cross covariance can be written as
\begin{align}
    \label{ec:var_x12b}
    P_{x_1x_2}(k+1) &= \alpha_1 P_{x_1}(k) \gamma_2^{\top} + \alpha_1 P_{x_1x_2}(k) \alpha_2^{\top},
\end{align}
where it is possible to see that $P_{x_1x_2}(k)$ converges only if $\rho(\alpha_2)<1$. Note that $P_{x_1x_2}(k)$ convergence requires $P_{x_1}(k)$ to also converge, which is guaranteed with $\rho(\alpha_1)<1$, $\rho((\alpha_1 \otimes \alpha_1) +\delta_1)<1$ and $M_{b_1}(1)=0$. 
For $P_{x_2}$, the diagonal structure of $P_{\Theta_N}$ allow us to write the variance of the state for the second follower as
\begin{align}
    \notag
    P_{x_2}(k+1) =& \Omega_{2}(k) +R_{2}(k)+\Lambda_{2}(k)+ \alpha_2 P_{x_{2}}(k) \alpha_2^{\top} \\
    &+B_2 \left[ P_{\theta_2} \odot ({C_{v_2}}\ P_{x_2}(k) {{C_{v_2}}}^\top) \right] B_2^\top. \notag 
\end{align}
We can see that the convergence of $P_{x_2}(k)$ requires that  $\Omega_{2}(k)$,$R_{2}(k)$ and $\Lambda_{2}(k)$ converge. $\Omega_{2}(k)$ converge if and only if $ \rho(\alpha_1)<1$, $\rho(\alpha_2)<1$, $M_{b_1}(1)=0$ and $M_{b_2}(1)=0$. $R_{2}(k)$ converge if and only if 
$P_{x_1}$ does, which is analyzed above. $\Lambda_{2}(k)$ converge if and only if $P_{x_1x_2}(k)$ converges, which is also analyzed above. Thus, it is not difficult to see that, for $P_{x_2}(k)$ to converge, the additionally condition $\rho((\alpha_2 \otimes \alpha_2) +\delta_2)<1$ must be included. 

For any $N>1$ we have that the cross covariance between the last vehicle and the rest of the platoon is given as
\begin{align}
    \label{ec:var_xN_N1b}
    P_{{\bf{x_{N-1}}}x_{N}}(k+1) &= \mathbf{A}_{\bf{N-1}} P_{\bf{x_{N-1}}}(k) \Psi_{\bf{N}}^{\top} \notag \\
    &\quad \quad + \mathbf{A}_{\bf{N-1}} P_{{\bf{x_{N-1}}}x_{N}}(k) \alpha_N^{\top} .
\end{align}
Assuming that $P_{\bf{x_{N-1}}}(k)$ converges, then we see that we require  $\rho(\mathbf{A}_{\bf{N-1}} \otimes \alpha_N)<1$ for $P_{{\bf{x_{N-1}}}x_{N}}(k)$ to converge. Given the structure of $\mathbf{A}_{\bf{N-1}}$ we notice that $\rho(\mathbf{A}_{\bf{N-1}} \otimes \alpha_N)<1 \Leftrightarrow  \max(\rho(\mathbf{A}_{\bf{N-1}}),\rho(\alpha_N))<1$. This is
implies that $\rho(\alpha_N)<1$.
The covariance matrix of the $N$-th vehicle can then be written as
\begin{align}
\notag
    P_{x_{N}}(k+1) &= \Omega_{N}(k)+ R_{N}(k) +\Lambda_{N}(k)+ \alpha_N P_{x_{N}}(k) \alpha_N^{\top} \\
    + B_N& \left[ P_{\theta_N} \odot (C_{v_N} P_{x_{N}}(k)\ C_{v_N}^{\top}) \right] B_N^{\top} 
        \label{convP}
\end{align}
$\Lambda_{N}(k)$ converge if $P_{{\bf{x_{N-1}}}x_{N}}(k)$ does, which requires $\rho(\alpha_N)<1$. If additionally we require
$M_{b_N}(1)=0$, then we ensure $\Omega_{N}(k)$ to converge. $R_{N}(k)$ converge given that $P_{\bf{x_{N-1}}}(k)$ does. Thus, it is not difficult to see from \eqref{convP} that additionally requiring 
that $\rho(\alpha_{N} \otimes \alpha_{N}+\delta_{N})<1$ ensures the convergence of $P_{x_{N}}(k)$.

Hence, applying these individual conditions recursively yields the second statement of Theorem \ref{teo:ind_mss_first_agent}.  \qed 



\end{document}